\documentclass[11pt,a4paper]{article}
\usepackage{geometry}
\usepackage{graphicx}
\usepackage{cite} 

\usepackage[utf8]{inputenc}
\usepackage[T1]{fontenc}
\usepackage{amsmath, amssymb, amsthm, mathtools}
\usepackage{slashed}
\usepackage{physics}
\usepackage{caption}    
\usepackage{bm}
\usepackage{booktabs}
\usepackage{tikz}
\usepackage{float}
\usepackage{pgfplots}
\pgfplotsset{compat=1.18}
\usepackage{xcolor}
\usepackage{hyperref} 
\usepackage{algorithmic}
\usepackage{array}
\usepackage{orcidlink}
\usepackage{stfloats}
\usepackage[caption=false,font=footnotesize]{subfig}
\usetikzlibrary{arrows, positioning, shapes}
\usetikzlibrary{arrows.meta, positioning, shapes.geometric}
\usetikzlibrary{positioning, arrows.meta, shapes.geometric, calc}
\definecolor{darkblue}{rgb}{0,0,0.6}
\newtheorem{remark}{Remark}
\newcommand{\thetaBar}{\bar{\theta}}
\newcommand{\alphaDot}{\dot{\alpha}}
\newcommand{\betaDot}{\dot{\beta}}

\newcommand{\V}{\mathcal{V}}
\newcommand{\cV}{\mathcal{V}}

\newcommand{\hatxi}{\hat{\xi}}

\newcommand{\Vtot}{V_{\mathrm{total}}}

\newcommand{\KKLT}{\texorpdfstring{\textsc{kklt}}{KKLT}}
\newcommand{\chiBar}{\bar{\chi}}
\newcommand{\lambdaBar}{\bar{\lambda}}

\newcommand{\Nequals}[1]{\mathcal{N}{=}#1}

\newcommand{\MF}{\mathcal{F}}

\newcommand{\CB}{\mathcal{M}_{\mathrm{CB}}}
\newcommand{\Mpl}{M_{\mathrm{Pl}}}
\newcommand{\myvev}[1]{\langle #1 \rangle}

\title{Supersymmetry, Supergravity and Non--Perturbative Dynamics of Gauge Theories} \author{} 

\author{%
  Tetiana Obikhod\\
  \it Institute for Nuclear Research,\\
  \it National Academy of Sciences of Ukraine,
  \it Kyiv, Ukraine\\
  \texttt{obikhod@kinr.kiev.ua}
}
\date{}

\begin{document}

\maketitle

\begin{abstract}
This article presents a comprehensive review of supersymmetry, supergravity, and
the non-perturbative dynamics of gauge theories, tracing a coherent path from the
algebraic structure of the supersymmetry algebra to moduli stabilisation and
de~Sitter vacua in string theory. We discuss representations of the supersymmetry
algebra, the superspace formalism, and basic supersymmetric models including the
Wess--Zumino model and $\mathcal{N}=1$ supersymmetric Yang--Mills theory. The
non-perturbative dynamics of $\mathcal{N}=2$ gauge theories is analysed in detail
through the Seiberg--Witten solution: the construction of the Seiberg--Witten
curve, the prepotential, the Picard--Fuchs system, the spectrum of BPS states, and
the mechanism of confinement via monopole condensation. The transition from global
supersymmetry to $\mathcal{N}=1$ supergravity is carried out in three steps,
revealing how the K\"{a}hler potential $K$ and superpotential $W$ completely
determine all five sectors of the Lagrangian and how the scalar potential acquires
its characteristic exponential prefactor and gravitationally induced negative
contribution. Applications to superstring theory are discussed, including D-brane
gauge theories, the AdS/CFT correspondence, geometric engineering of the
Seiberg--Witten solution, and the mechanisms by which $\mathcal{N}=4$ supersymmetry
is reduced to phenomenologically viable $\mathcal{N}=1$ models. Finally, we analyse
the KKLT moduli stabilisation mechanism in detail, including the effect of
$\alpha'^3$ corrections to the K\"{a}hler potential. We identify three distinct
regimes of the resulting scalar potential --- classical KKLT, corrected KKLT with
a shifted AdS minimum, and a runaway regime corresponding to the Large Volume
Scenario --- and determine the critical correction parameter $\hat{\xi}_c$ that
separates controlled de~Sitter vacua from decompactification. The tension between
this structure and the de~Sitter swampland conjecture is discussed.
\end{abstract}

\tableofcontents

\section{Introduction}
Supersymmetry occupies a singular place in theoretical physics. Unlike most symmetry principles, which impose constraints on the matter content or interactions of a theory while leaving its ultraviolet behaviour essentially unchanged, supersymmetry restructures the very way quantum corrections accumulate. The cancellations between bosonic and fermionic loops that it enforces are not accidents of perturbation theory but consequences of an algebra that extends the Poincar\'e group by fermionic generators $Q_\alpha$ and $\bar{Q}_{\dot\alpha}$ satisfying $\{Q_\alpha, \bar{Q}_{\dot\alpha}\} = 2\sigma^\mu_{\alpha\dot\alpha}P_\mu$. It is this algebraic rigidity that makes supersymmetric theories tractable far beyond the reach of ordinary perturbative methods, and that has sustained interest in the subject over five decades.

The clearest demonstration of this tractability came with the work of Seiberg and Witten~\cite{Seiberg1994a,Seiberg1994b} on four-dimensional $\mathcal{N}=2$ supersymmetric Yang--Mills theory. What they showed was unexpected in its directness: the full non-perturbative dynamics of the theory, at all energy scales and for all values of the coupling, could be encoded in the geometry of a single algebraic curve. The low-energy effective action on the Coulomb branch --- where $SU(2)$ is spontaneously broken to $U(1)$ and the physics is that of a single massless vector multiplet --- is completely determined by a holomorphic prepotential $\mathcal{F}(a)$, and this prepotential is in turn determined by the period integrals $a = \oint_\alpha \lambda_{\mathrm{SW}}$ and $a_D = \oint_\beta \lambda_{\mathrm{SW}}$ of the Seiberg--Witten differential $\lambda_{\mathrm{SW}} = x\,dx/y$ over the two independent cycles of an elliptic curve $y^2 = (x-e_1)(x-e_2)(x-e_3)$. The modular parameter $u = \mathrm{Tr}\,\varphi^2$ of this curve is precisely the gauge-invariant coordinate on the Coulomb branch, and as $u$ varies the complex structure of the torus deforms continuously, carrying with it all the information about the quantum vacuum.

This geometric encoding of dynamics has several striking consequences. The BPS states of the theory --- monopoles, dyons, and electrically charged particles, whose masses saturate the bound $M = |n_e a + n_m a_D|$ --- can be read off directly from the period integrals. Electric--magnetic duality becomes the statement that the monodromy group of the periods over the $u$-plane is $SL(2,\mathbb{Z})$, the group of automorphisms of the charge lattice $\mathbb{Z}^2$. And confinement, long understood as the magnetic dual of the Higgs mechanism, receives its first exact field-theoretic proof: a soft mass term $\mu\,\mathrm{Tr}\,\varphi^2$ selects vacua near the singular points $u = \pm\Lambda^2$ where a monopole or dyon becomes massless, condenses, and produces colour-confining electric flux tubes.

While powerful, the Seiberg--Witten solution describes a theory without gravity. Connecting it to a complete physical framework requires two further steps~\cite{Polchinski,WessBagger1992}. The first is the localisation of supersymmetry: when the parameter $\epsilon_\alpha$ is allowed to depend on the spacetime point, the algebra demands the introduction of a spin-$3/2$ gravitino $\psi_{\mu\alpha}$ and a spin-$2$ graviton $g_{\mu\nu}$, giving $\mathcal{N}=1$ supergravity. The dynamics of the resulting theory is governed by the K\"ahler potential $K$ and superpotential $W$, two holomorphic functions that together determine all five sectors of the Lagrangian --- kinetic terms, gauge couplings, scalar potential, Yukawa interactions, and the four-fermion contact term. The transition from the globally supersymmetric to the locally supersymmetric theory is not merely the addition of gravitational interactions: it modifies the scalar potential by an exponential factor $e^{K/M_{\mathrm{Pl}}^2}$ and introduces a negative gravitational contribution $-3|W|^2/M_{\mathrm{Pl}}^2$ that allows the vacuum energy to take any sign.

The second step is the embedding of supergravity into superstring theory, where extra dimensions, D-brane dynamics, and flux backgrounds provide a geometric origin for the abstract field-theoretic structures. A stack of $N$ coincident D3-branes carries $\mathcal{N}=4$ super Yang--Mills on its worldvolume; placing the branes at orbifold singularities or threading the compactification with quantised fluxes reduces the supersymmetry to $\mathcal{N}=2$ or $\mathcal{N}=1$ and generates the gauge content and superpotential couplings needed for phenomenology. The Seiberg--Witten curve itself re-emerges in this context as the geometry of the brane configuration, and the period integrals that encode the gauge dynamics appear as integrals of the holomorphic three-form $\Omega_3$ over two-cycles in the internal space.

A central challenge at this interface is the stabilisation of moduli --- the massless scalar fields that parametrise the size and shape of the extra dimensions. Without stabilisation, these fields would mediate unobserved long-range forces and cause couplings to vary in time. The KKLT construction addresses this by combining a flux-generated superpotential $W_0$ with a non-perturbative gaugino-condensation term $Ae^{-aT}$ to fix all moduli at a supersymmetric AdS minimum, and then uplifting to a metastable de~Sitter vacuum through the positive energy contribution of $\overline{\mathrm{D3}}$-branes. The self-consistency of this construction, however, is sensitive to higher-derivative corrections to the K\"ahler potential. In particular, the $\alpha'^3$ correction breaks the no-scale identity $K^{T\bar{T}}(\partial_T K)(\partial_{\bar{T}}K)=3$ and shifts the AdS minimum to larger volume; we analyse in detail when this correction preserves the minimum and when it drives a runaway, identifying three distinct regimes of the potential that connect classical KKLT to the Large Volume Scenario and to outright decompactification.

The purpose of this work is to trace how supersymmetry encodes dynamical questions in geometric structures, and how this correspondence persists under localisation and string compactification. The exposition proceeds as follows. Section~\ref{sec:SYM} establishes the $\mathcal{N}=1$ superspace formalism and derives the component Lagrangian, emphasising how the K\"ahler potential and superpotential fix all interaction sectors. Section~\ref{sec:SW} analyses the exact non-perturbative dynamics of $\mathcal{N}=2$ gauge theories through the Seiberg--Witten solution, detailing the geometry of the Coulomb branch, the Picard--Fuchs system, BPS spectra, and monopole-driven confinement. Section~\ref{sec:SUGRA} carries out the localisation of supersymmetry, demonstrating how curved superspace deforms the scalar potential and couples the geometry of the scalar manifold to spacetime. In Section~\ref{sec:strings}, both constructions are embedded into string theory, where D-brane configurations and flux backgrounds provide a concrete geometric origin for the abstract field-theoretic data. Section~\ref{sec:moduli} applies this framework to the KKLT moduli-stabilisation mechanism, systematically analysing the impact of $\alpha'^3$ corrections, mapping three distinct vacuum regimes, and confronting the resulting parameter space with the de~Sitter swampland conjecture. Section~\ref{sec:conclusion} summarises the findings and outlines directions for further investigation.
%%%%%%%%%%%%%%%%%%%%%%%%%%%%%%%%%%%%%%%

\section{Supersymmetric Yang-Mills Theory}\label{sec:SYM}

\subsection{Motivation for Supersymmetry}

Supersymmetry unifies matter and force carriers by extending the Poincar\'e 
algebra with fermionic generators $Q_\alpha$ and $\bar Q_{\dot\alpha}$, 
so that each boson acquires a fermionic superpartner and vice 
versa\cite{WessBagger1992,WeinbergQFT3}.
Beyond this aesthetic unification, the symmetry stabilises scalar masses against 
large radiative corrections — the hierarchy problem — and emerges 
as an indispensable ingredient in both supergravity and superstring theory.
The algebra takes the form
\begin{equation}
\{ Q_\alpha , \bar Q_{\dot{\alpha}} \} =
2 \sigma^\mu_{\alpha \dot{\alpha}} P_\mu ,
\end{equation}
where $P_\mu$ is the generator of spacetime translations and $\sigma^\mu$ are the Pauli matrices.
The remaining anticommutators vanish

\begin{equation} \{Q_\alpha,Q_\beta\}=0, \qquad \{\bar Q_{\dot\alpha},\bar Q_{\dot\beta}\}=0 \end{equation}
The Lorentz generators act on the supercharges as
\begin{equation} [M_{\mu\nu},Q_\alpha] = (\sigma_{\mu\nu})_\alpha{}^{\beta} Q_\beta \end{equation}
Bosonic and fermionic fields are grouped in representations called \textit{supermultiplets}. 
To obtain the \textit{supermultiplets}, one expands the corresponding superfields in powers of the Grassmann coordinates $\theta$ and $\bar{\theta}$ of superspace \cite{WessBagger1992,GatesSuperspace}. This expansion allows the superfield to be written explicitly in terms of ordinary spacetime fields, which represent the physical degrees of freedom. As a result, the bosonic and fermionic components naturally appear as elements of the same supermultiplet, together with auxiliary fields that ensure the off-shell closure of the supersymmetry algebra. Superspace coordinates are defined as the extended set:
\begin{equation}
    z^M = (x^\mu, \theta_\alpha, \thetaBar^{\alphaDot}),
    \label{eq:superspace_coords}
\end{equation}
where $\theta_\alpha$ and $\thetaBar_{\alphaDot}$ are independent Grassmann variables satisfying the anticommutation relations:
\begin{equation}
    \{\theta_\alpha, \theta_\beta\} = 0, \quad \{\thetaBar_{\alphaDot}, \thetaBar_{\betaDot}\} = 0, \quad \{\theta_\alpha, \thetaBar_{\alphaDot}\} = 0.
\end{equation}

A general scalar superfield admits the expansion:
\begin{align}
    \Phi(x,\theta,\thetaBar) &= A(x) + \theta\psi(x) + \thetaBar\chiBar(x) \nonumber \\
    &+ \theta\theta F(x) + \thetaBar\thetaBar G(x) + \theta\sigma^\mu\thetaBar v_\mu(x) \nonumber \\
    &+ \theta\theta\thetaBar\lambdaBar(x) + \thetaBar\thetaBar\theta\rho(x) + \theta\theta\thetaBar\thetaBar D(x),
    \label{eq:superfield_expansion}
\end{align}
where $A, F, G, D$ are complex scalar fields, $\psi, \chi, \lambda, \rho$ are Weyl fermions, and $v_\mu$ is a vector field.

In four-dimensional supersymmetric theories, the simplest and most commonly studied representations are the minimal $\mathcal{N}=1$ supermultiplets. These multiplets contain the smallest possible number of fields required to realize supersymmetry in four dimensions. Each multiplet combines bosonic and fermionic degrees of freedom in such a way that their numbers are equal, reflecting the fundamental supersymmetric relation between particles of integer and half-integer spin. The minimal $\mathcal{N}=1$ framework therefore provides the basic building blocks from which more complex supersymmetric models are constructed. So, the simplest representations are the chiral multiplet and vector multiplet with chiral multiplet presented by
\begin{equation} (\phi,\psi,F) \end{equation}
where $\phi$ is a complex scalar field, $\psi$ is a Weyl fermion and $F$ is an auxiliary field.
The vector multiplet is
\begin{equation} (A_\mu,\lambda,D) \end{equation}
where $A_\mu$ is the gauge field, $\lambda$ is the gaugino and $D$ is an auxiliary scalar field. In supersymmetric models, chiral multiplets are used to describe {\it{matter fields}}. In particular, the quarks and leptons of the Standard Model, together with their scalar superpartners (squarks and sleptons), are naturally incorporated into chiral multiplets. The vector multiplet is responsible for describing {\it{gauge interactions}} in supersymmetric theories. The field $A_\mu$ plays the role of the usual gauge boson that mediates interactions, while its fermionic superpartner $\lambda$ completes the multiplet required by supersymmetry. The auxiliary field $D$ does not represent a physical propagating degree of freedom; instead, it is introduced to ensure the proper realization of the supersymmetry algebra at the level of component fields.
\subsection{Kinetic Terms of Supersymmetric Multiplets}

The dynamics of the fields contained in supersymmetric multiplets are described by their kinetic terms in the Lagrangian. These terms determine how the component fields propagate in spacetime. For the chiral multiplet, the kinetic part of the Lagrangian includes the usual kinetic term for the scalar field, the fermionic kinetic term for the Weyl spinor, and a quadratic term for the auxiliary field. It can be written as
\begin{equation}
\mathcal{L}_{\text{chiral}} =
|\partial_\mu \phi|^2
+ i \bar{\psi}\bar{\sigma}^{\mu}\partial_{\mu}\psi
+ |F|^2 .
\end{equation}
Here the first term describes the propagation of the complex scalar field $\phi$, the second term corresponds to the kinetic term of the fermionic field $\psi$, and the last term involves the auxiliary field $F$, which does not represent a physical propagating degree of freedom but is required for the off--shell formulation of supersymmetry.

In a similar way, the kinetic terms for the vector multiplet describe the dynamics of the gauge field and its fermionic superpartner. The supersymmetric Yang-Mills Lagrangian for N = 1 gauge theory \cite{WessBagger1992,MartinSUSY} takes the form
\begin{equation}
\mathcal{L}_{\text{vector}} =
-\frac{1}{4} F_{\mu\nu}^aF^{a\mu\nu}
+ i\bar{\lambda}^a\bar{\sigma}^{\mu}D_{\mu}\lambda^a
+ \frac{1}{2}D^aD^a .
\end{equation}
The first term is the familiar kinetic term for the gauge field $A_\mu^a$, expressed through the field strength tensor $F_{\mu\nu}^a = \partial_\mu A_\nu^a - \partial_\nu A_\mu^a + g f^{abc} A_\mu^b A_\nu^c$. The second term represents the kinetic contribution of the gaugino field $\lambda^a$, while the final term involves the auxiliary field $D^a$, an auxiliary field necessary for off-shell closure of the supersymmetry algebra.

Supersymmetry transformations take the schematic form:
\begin{align}
    \delta_\epsilon A_\mu^a &= i\epsilon \sigma_\mu \bar{\lambda}^a - i\bar{\epsilon} \bar{\sigma}_\mu \lambda^a, \\
    \delta_\epsilon \lambda^a_\alpha &= \sigma^{\mu\nu}_{\alpha}{}^{\beta} \epsilon_\beta F_{\mu\nu}^a + i\epsilon_\alpha D^a, \\
    \delta_\epsilon D^a &= -\epsilon \sigma^\mu D_\mu \bar{\lambda}^a + \bar{\epsilon} \bar{\sigma}^\mu D_\mu \lambda^a,
    \label{eq:SUSY_transformations}
\end{align}
where $\epsilon_\alpha$ is the infinitesimal Grassmann-valued supersymmetry parameter, and $\sigma^{\mu\nu} = \frac{i}{4}(\sigma^\mu \bar{\sigma}^\nu - \sigma^\nu \bar{\sigma}^\mu)$. 

The supersymmetric covariant derivatives $D_\alpha$ which enforce consistency with both general covariance and supersymmetry, are given by:
\begin{align}
    D_\alpha &= \frac{\partial}{\partial\theta^\alpha} + i\sigma^\mu_{\alpha\alphaDot}\thetaBar^{\alphaDot}\partial_\mu, \\
    \bar{D}_{\alphaDot} &= -\frac{\partial}{\partial\thetaBar^{\alphaDot}} - i\theta^\alpha\sigma^\mu_{\alpha\alphaDot}\partial_\mu,
\end{align}
satisfying the anticommutation relations:
\begin{equation}
    \{D_\alpha, \bar{D}_{\alphaDot}\} = -2i\sigma^\mu_{\alpha\alphaDot}\partial_\mu, \quad \{D_\alpha, D_\beta\} = 0, \quad \{\bar{D}_{\alphaDot}, \bar{D}_{\betaDot}\} = 0.
\end{equation}

{\bf{Chiral superfields}} satisfy the constraint $\bar{D}_{\alphaDot}\Phi = 0$ and depend on the chiral coordinates $y^\mu = x^\mu + i\theta\sigma^\mu\thetaBar$:
\begin{equation}
    \Phi(y,\theta) = \phi(y) + \sqrt{2}\theta\psi(y) + \theta\theta F(y).
\end{equation}
In terms of the original coordinates, this expands to:
\begin{align}
    \Phi(x,\theta,\thetaBar) &= \phi(x) + i\theta\sigma^\mu\thetaBar\partial_\mu\phi(x) + \frac{1}{4}\theta\theta\thetaBar\thetaBar\Box\phi(x) \nonumber \\
    &+ \sqrt{2}\theta\psi(x) - \frac{i}{\sqrt{2}}\theta\theta\partial_\mu\psi(x)\sigma^\mu\thetaBar + \theta\theta F(x).
\end{align}
{\bf {The gauge field}} is embedded in a real vector superfield $V = V^a T^a$ transforming under gauge transformations as:
\begin{equation}
    e^{2gV} \rightarrow e^{-i\Lambda^\dagger} e^{2gV} e^{i\Lambda},
\end{equation}
where $\Lambda$ is a chiral superfield parameter. In terms of the original coordinates, the vector superfield expands as:
\begin{equation}
    V^a(x,\theta,\thetaBar) = -\theta\sigma^\mu\thetaBar A_\mu^a(x) + i\theta\theta\thetaBar\bar{\lambda}^a(x) - i\thetaBar\thetaBar\theta\lambda^a(x) + \frac{1}{2}\theta\theta\thetaBar\thetaBar D^a(x).
\end{equation}

The gauge-covariant field strength superfield is defined as:
\begin{equation}
    W_\alpha^a = -\frac{1}{4}\bar{D}\bar{D} \left(e^{-2gV} D_\alpha e^{2gV}\right)^a,
\end{equation}
which satisfies the chirality constraint $\bar{D}_{\alphaDot} W_\alpha = 0$ and transforms covariantly under gauge transformations.

{\bf {Supersymmetric actions}} can be written as superspace integrals:
\begin{equation}
    S = \int d^4x d^4\theta \, K(\Phi_i, \Phi_i^\dagger e^{2gV}) + \left[\int d^4x d^2\theta \, W(\Phi_i) + \text{h.c.}\right],
    \label{eq:action_global}
\end{equation}
where $K$ is the Kähler potential (a real function) and $W$ is the holomorphic superpotential.

The gauge kinetic term is given by:
\begin{equation}
    S_{\text{gauge}} = \frac{1}{4g^2}\int d^4x d^2\theta \, \Tr(W^\alpha W_\alpha) + \text{h.c.}
    \label{eq:gauge_kinetic}
\end{equation}

\subsection{Interactions in Supersymmetric Theories}
Interactions between fields in supersymmetric theories arise in a structured and constrained way due to supersymmetry. After performing the Grassmann integration in~\eqref{eq:action_global} and eliminating auxiliary fields via their equations of motion, the component Lagrangian for a general $N=1$ supersymmetric gauge theory with chiral matter becomes:

\begin{align}
    \mathcal{L} &= -\frac{1}{4}F_{\mu\nu}^a F^{a\mu\nu} + i\bar{\lambda}^a \bar{\sigma}^\mu D_\mu \lambda^a \nonumber \\
    &+ (D_\mu\phi_i)^\dagger (D^\mu\phi_i) + i\bar{\psi}_i \bar{\sigma}^\mu D_\mu \psi_i \nonumber \\
    &- \sqrt{2}g\left(\phi_i^\dagger T^a_{ij} \psi_j \lambda^a + \bar{\lambda}^a \bar{\psi}_i T^a_{ij} \phi_j\right) \nonumber \\
    &- V(\phi, \phi^\dagger) - \frac{1}{2}\left(\frac{\partial^2 W}{\partial\phi_i\partial\phi_j}\psi_i\psi_j + \text{h.c.}\right),
    \label{eq:component_lagrangian}
\end{align}
where the scalar potential is given by:
\begin{equation}
    V(\phi, \phi^\dagger) = \sum_i \left|\frac{\partial W}{\partial\phi_i}\right|^2 + \frac{g^2}{2}\sum_a \left(\phi_i^\dagger T^a_{ij} \phi_j\right)^2.
    \label{eq:scalar_potential}
\end{equation}
In $\mathcal{N}=1$ models, the main sources of interactions are the superpotential, 
Yukawa-type couplings, and gauge interactions. The component Lagrangian 
\eqref{eq:component_lagrangian} organizes these into three fundamental sectors:

\begin{enumerate}
    \item \textbf{Gauge sector:} The gauge kinetic term 
    $-\frac{1}{4}F_{\mu\nu}^a F^{a\mu\nu}$ describes the dynamics of gauge fields, 
    while $i\bar{\lambda}^a\bar{\sigma}^\mu D_\mu \lambda^a$ provides the 
    covariant kinetic term for their fermionic superpartners, the gauginos. 
    These arise from the coupling between chiral multiplets and vector multiplets 
    and describe how matter fields interact with gauge fields and gauginos. 
    The gauge structure ensures that both bosonic and fermionic components 
    participate consistently in the same interaction framework.

    \item \textbf{Matter kinetic sector:} The terms 
    $(D_\mu\phi_i)^\dagger (D^\mu\phi_i)$ and 
    $i\bar{\psi}_i\bar{\sigma}^\mu D_\mu \psi_i$ provide the covariant kinetic 
    terms for scalar and fermionic components of chiral multiplets, respectively.

    \item \textbf{Interaction sector:} This contains two distinct types of 
    couplings:
    \begin{itemize}
        \item \textit{Gauge Yukawa couplings:} $-\sqrt{2}g\left(\phi_i^\dagger 
        T^a_{ij}\psi_j\lambda^a + \text{h.c.}\right)$ describe direct 
        interactions between matter scalars, matter fermions, and gauginos.
        
        \item \textit{Superpotential Yukawa couplings:} $-\frac{1}{2}\left(
        \frac{\partial^2 W}{\partial\phi_i\partial\phi_j}\psi_i\psi_j + 
        \text{h.c.}\right)$ emerge from the superpotential $W(\Phi)$, which 
        is a holomorphic function of the chiral superfields. The superpotential 
        plays a central role in determining the scalar potential 
        $V(\phi,\phi^\dagger)$ and the interaction structure of the theory. 
        Through the superpotential, scalar fields and fermions belonging to 
        chiral multiplets interact with each other in a way that preserves 
        supersymmetry. Yukawa interactions naturally appear when the 
        superpotential is expanded in terms of component fields. These 
        interactions couple scalar fields to fermions and are responsible for 
        generating fermion masses and interaction vertices after symmetry 
        breaking.
    \end{itemize}
\end{enumerate}

The scalar potential $V(\phi,\phi^\dagger)$ occupies a central role in determining the physical consequences of the theory, as it governs the vacuum structure and the patterns of spontaneous symmetry breaking. This potential receives contributions from two distinct sources: the $F$-terms, which arise from the superpotential and encode the self-interactions of chiral superfields, and the $D$-terms, which originate from the gauge sector and enforce the local symmetry constraints.

The minimization of this potential identifies the vacuum state of the theory, revealing whether the ground state preserves or spontaneously breaks the underlying symmetries. When scalar fields acquire nonzero vacuum expectation values at the minimum, gauge symmetries may be broken, thereby generating masses for gauge bosons and their superpartners through the Higgs and super-Higgs mechanisms. Meanwhile, the $F$-term contributions determine the soft-breaking masses and trilinear couplings that split the degeneracy within supermultiplets.

In this manner, the scalar potential serves as the bridge between the formal supersymmetric structure and observable phenomenology. It transforms the abstract field content into concrete predictions for mass spectra, coupling constants, and the hierarchy of symmetry breaking scales that characterize the low-energy effective theory.

\subsection{The Wess-Zumino Model}

The Wess-Zumino model is one of the most pedagogically valuable examples of a four-dimensional supersymmetric quantum field theory \cite{WessZumino,SeibergHolomorphy}. It is a rigorous framework for discussing supersymmetry breaking, renormalization, and the interplay between bosonic and fermionic degrees of freedom. Although this theory is seemingly very simple, the Wess-Zumino model is phenomenologically rich and has played a crucial role in the development of supersymmetric theories.

The classical action of the Wess-Zumino model is constructed from a single chiral superfield $\Phi$, whose component expansion in four-dimensional $\mathcal{N}=1$ superspace reads:
\begin{equation}
\Phi(x, \theta, \bar{\theta}) = \phi(x) + \sqrt{2}\theta \psi(x) + \theta^2 F(x),
\end{equation}
where $\phi(x)$ denotes a complex scalar field, $\psi(x)$ is a two-component Weyl fermion, and $F(x)$ represents the auxiliary scalar field. The Lagrangian density for the free theory is given by:
\begin{equation}
\mathcal{L}_0 = \int d^2\theta \, d^2\bar{\theta} \, \Phi^\dagger \Phi.
\end{equation}
This kinetic term, when expanded in components, yields the standard kinetic energies for the scalar and fermion fields, along with the kinetic term for the auxiliary field $F$. The introduction of interactions proceeds through a superpotential $W(\Phi)$, which must be a holomorphic function of the chiral superfield. For the most commonly studied case, we consider the polynomial superpotential:
\begin{equation}
W(\Phi) = m\Phi + \frac{\lambda}{3}\Phi^3,
\end{equation}
where $m$ is the mass parameter and $\lambda$ denotes the coupling constant. The full Lagrangian is then expressed as:
\begin{equation}
\mathcal{L} = \int d^2\theta \, d^2\bar{\theta} \, \Phi^\dagger \Phi + \left(\int d^2\theta \, W(\Phi) + \text{h.c.}\right).
\end{equation}

One of the most interesting aspects of the Wess-Zumino model is its perturbative renormalizability in four dimensions. This is in sharp contrast with many other quantum field theories and is a result of the non-renormalisation theorems satisfied by supersymmetric theories. The renormalisation of the superpotential is absent beyond one-loop order due to holomorphicity and supersymmetric Ward identities. The anomalous dimensions of the superfields vanish in the leading approximation, and quantum effects are completely encoded in wave function renormalisation of the fields. This has made the Wess-Zumino model an invaluable tool in studying renormalisation group flows in supersymmetric theories.

The spontaneous breaking of supersymmetry in the Wess-Zumino model occurs through the non-vanishing vacuum expectation value (VEV) of the auxiliary field $F$. At the classical level, the scalar potential derived from the superpotential is:
\begin{equation}
V(\phi, \phi^\dagger) = |F|^2 + |G|^2 = \left|\frac{\partial W}{\partial \phi}\right|^2 + \left|\frac{\partial W^\dagger}{\partial \phi^\dagger}\right|^2,
\end{equation}
where $G = \partial W^\dagger/\partial \phi^\dagger$ is the auxiliary field equation of motion. For the linear case ($W = m\Phi$), the potential vanishes identically, indicating unbroken supersymmetry. However, for the cubic case with $W = m\Phi + (\lambda/3)\Phi^3$, the potential exhibits a non-trivial structure that can lead to supersymmetry breaking at the quantum level through loop corrections to the effective potential.

Moreover, the Wess-Zumino model provides an essential framework for understanding the concept of holomorphy and non-renormalization theorems in supersymmetric quantum field theory. The beta functions and anomalous dimensions can be calculated exactly using superspace methods and the operator product expansion, giving us exact results that go beyond what we can obtain with perturbation theory, unlike what happens in non-supersymmetric quantum field theory. The implications of the exact results are very profound, not only for understanding the structure of supersymmetric quantum field theory at weak and strong coupling, but also for encouraging the search for other exactly solvable supersymmetric quantum field theories that are relevant to physics beyond the Standard Model.

 The enlargement of the supersymmetry algebra from N=1  to higher N  imposes powerful constraints: the scalar field manifold becomes rigidly fixed to symmetric spaces, non-renormalization theorems strengthen, and ultraviolet divergences progressively diminish. In the maximally supersymmetric case of N=4  super Yang-Mills, these constraints culminate in exact conformal invariance and finiteness, rendering the theory an invaluable theoretical laboratory for exploring non-perturbative phenomena and strong-weak dualities. The introduction of multiple supercharges addresses a fundamental tension in constructing realistic unified theories: the need to accommodate both gravitational and gauge interactions within a consistent quantum framework. Extended supersymmetry provides the requisite mathematical structure to achieve this unification, as the additional fermionic symmetries enforce relations between bosonic and fermionic degrees of freedom that prove essential for anomaly cancellation and vacuum stability. Furthermore, the enhanced symmetry restricts the available counterterms, thereby improving the ultraviolet behavior and offering a viable path toward ultraviolet completion in supergravity and string-theoretic constructions. Extended supersymmetry with multiple supercharges is presented by 

\begin{equation} Q^I_\alpha , \qquad I=1,...,N \end{equation}

The algebra becomes

\begin{equation} \{Q^I_\alpha,\bar Q^J_{\dot\beta}\} = 2\delta^{IJ}\sigma^\mu_{\alpha\dot\beta}P_\mu \end{equation}

Extended supersymmetry strongly constrains quantum corrections and often leads to improved ultraviolet behaviour.

%%%%%%%%%%%%%%%%%%%%%%%%%%%%%%%%%%%%%%%%%%%%%%%%%%%%%%%%

\section{Non--Perturbative Dynamics and Seiberg--Witten Theory}\label{sec:SW}

The study of strongly coupled quantum field theories remains one of the central challenges in modern theoretical physics, as conventional perturbative methods fail in this regime. A major breakthrough was achieved in the seminal work of Nathan Seiberg and Edward Witten, who developed what is now known as Seiberg–Witten theory \cite{Seiberg1994a, Seiberg1994b}. This framework provides an exact, non-perturbative solution to four-dimensional $N$ = 2 supersymmetric gauge theories.

Seiberg–Witten theory reveals that the low-energy effective dynamics of such theories can be described exactly in terms of a holomorphic prepotential and encoded geometrically by an associated algebraic curve, the Seiberg–Witten curve. This remarkable correspondence establishes a deep connection between quantum field theory and complex geometry.

One of the key achievements of the theory is the explicit demonstration of electric–magnetic duality, showing that strongly coupled regimes of the theory can be equivalently described by weakly coupled dual variables. In addition, it provides a concrete mechanism for confinement through monopole condensation and allows for the exact determination of the spectrum of stable BPS states.

Since the vacuum structure determines the low-energy physics entirely, we 
analyse it in detail, beginning with the superselection decomposition of the 
Hilbert space.

\subsection{Superselection Structure of the Vacuum Manifold
in Four-Dimensional \texorpdfstring{$\mathcal{N}=2$}{N=2} Supersymmetric Theories}

In quantum field
theory, the infinite number of degrees of freedom leads to a decomposition of
the full Hilbert space into mutually orthogonal superselection sectors, between
which no local observable possesses non-vanishing matrix elements. This structure
is absent in quantum mechanics, where tunneling amplitudes between degenerate
vacua remain finite. 

In quantum mechanics, a theory with
a discrete set of degenerate classical vacua---such as a scalar potential with
minima at $\varphi = \pm v$---admits a well-defined quantum ground state that
is a symmetric linear combination of the classical configurations,
\begin{equation}
    |\Omega\rangle = \frac{1}{\sqrt{2}}\bigl(|{+v}\rangle + |{-v}\rangle\bigr).
\end{equation}
The situation changes qualitatively in the thermodynamic limit of the quantum field
theory, where the system possesses an infinite number of degrees of freedom. To
transition from the vacuum at $\varphi = -v$ to that at $\varphi = +v$, the
field must be flipped throughout all of infinite space. The tunneling amplitude
is then proportional to the product of local factors, one for each spatial point,
\begin{equation}
    \mathcal{A}_{\rm tunnel} \;\sim\; \varepsilon^{\,V/a^3}
    \;\xrightarrow{\;V \to \infty\;}\; 0,
\end{equation}
where $\varepsilon < 1$ is the local tunneling probability, $V$ is the spatial
volume, and $a$ is a short-distance regulator. In the infinite-volume limit
this amplitude vanishes exactly, and the two vacua become completely
disconnected.

This phenomenon is captured by the notion of \emph{superselection sectors}:
the full Hilbert space $\mathcal{H}$ decomposes as a direct sum of orthogonal
subspaces,
\begin{equation}
    \mathcal{H} \;=\; \mathcal{H}_{+v} \oplus \mathcal{H}_{-v},
\end{equation}
with the defining property that every local observable $\mathcal{O}(x)$
has vanishing matrix elements between distinct sectors,
\begin{equation}
    \langle +v\,|\;\mathcal{O}(x)\;|\,-v\rangle \;=\; 0
    \quad \forall\; x,\; \forall\;\mathcal{O} \in \mathfrak{A}_{\rm loc}.
\end{equation}
This distinction shapes the geometry of the moduli space of vacua in $N=2$ supersymmetric field theories in four dimensions, where the Coulomb branch constitutes a continuous family of physically inequivalent superselection sectors, each supporting its own infrared $U(1)^r$ effective theory.

\subsection{Moduli space structure}
The physical significance of this structure becomes particularly transparent in
the context of four-dimensional $\mathcal{N}=2$ supersymmetric gauge theories, whose moduli spaces of vacua have been analyzed in depth since the seminal work of Seiberg and Witten~\cite{Seiberg1994a,Seiberg1994b}. In these theories, supersymmetry protects a continuous family of vacua against quantum corrections,
giving rise to the \emph{moduli space} $\mathcal{M}$ \cite{IntriligatorSeiberg}. The latter decomposes
into qualitatively distinct branches: the Coulomb
branch~$\mathcal{C}$, on which the scalar component of the $\mathcal{N}=2$
vector multiplet acquires a vacuum expectation value $\langle\varphi\rangle \neq 0$,
and the Higgs branch, on which the scalar hypermultiplet fields condense.

In $\mathcal{N}=2$ supersymmetric $SU(2)$ gauge theory, the vacuum structure is not isolated but forms a continuous moduli space parametrized by the gauge-invariant coordinate $u$. Each point $u \in \mathcal{C}$ labels a distinct superselection sector, corresponding to a physically inequivalent vacuum of the theory. Moving along $\mathcal{C}$ does not represent a
dynamical process within a single Hilbert space, but rather a parametric
labeling of the family of inequivalent vacua, each inaccessible from the
others by any local operation. The low-energy effective dynamics over this moduli space is elegantly encoded in a family of complex curves $\Sigma_u$, known as Seiberg--Witten curves. 
The vacuum space of this theory --- the \textbf{moduli space} $\mathcal{M}$ --- decomposes into qualitatively
distinct components. On the part of the distinct components ---\textbf{Coulomb branch},  the scalar component $\varphi$ of the
vector multiplet acquires a nonzero vacuum expectation value $\langle\varphi\rangle \neq 0$, which triggers
the symmetry breaking $\mathrm{SU}(2) \to \mathrm{U}(1)$. The gauge-invariant coordinate
\begin{equation}
    u = \mathrm{Tr}\,\varphi^2
\end{equation}
parametrizes this branch. Each point $u \in \mathcal{C}$ corresponds to a distinct superselection sector.
The low-energy physics at a generic point of $\mathcal{C}$ is described by a free Abelian $\mathrm{U}(1)$
theory with a single vector multiplet. The dynamical scale $\Lambda$ governs the strength of quantum effects.

The central object of the theory is the \textbf{Seiberg--Witten curve} --- a family of elliptic
curves parametrized by the point $u \in \mathcal{C}$ \cite{KlemmEtAl}. In Weierstrass form it is written as
\begin{equation}
    y^2 = (x - e_1)(x - e_2)(x - e_3),
\end{equation}
where the roots $e_i$ depend on $u$ and $\Lambda$. Each such curve is a \textbf{torus} --- a Riemann
surface of genus~1, that is, a complex one-dimensional manifold with a single ``handle.''
Thus, each point of the Coulomb branch corresponds to a specific member of this family of
elliptic curves, and motion along $\mathcal{C}$ is equivalent to an isocomplex deformation of
the complex structure of the torus.
On the torus there exist exactly two topologically non-trivial cycles --- the
\textbf{$\alpha$-cycle} and the \textbf{$\beta$-cycle} --- which cannot be contracted to a point.
These cycles determine the physically observable quantities of the theory through the
\textbf{geometric integrals} of the Seiberg--Witten differential $\lambda_{\mathrm{SW}}$ --- the meromorphic differential of the Seiberg--Witten theory defined explicitly as
\begin{equation}
    \lambda_{\mathrm{SW}} = \frac{x\, dx}{y},
\end{equation}
where $x$ and $y$ are related by the Seiberg--Witten curve
$y^2 = (x - e_1)(x - e_2)(x - e_3)$, with roots $e_i = e_i(u, \Lambda)$ depending
on the modular coordinate $u \in \mathcal{C}$ and the dynamical scale $\Lambda$.
The differential $\lambda_{\mathrm{SW}}$ is meromorphic on the torus $\Sigma_u$:
holomorphic at generic points, it develops poles at the branch points of the curve.
This analytic structure is not incidental --- it is precisely what allows
$\lambda_{\mathrm{SW}}$ to encode both perturbative and non-perturbative data of
the theory within a single geometric object.

\textbf{The central result of Seiberg--Witten theory} is that all infrared physical
observables are expressed as period integrals of $\lambda_{\mathrm{SW}}$ over the
two homologically independent cycles $\alpha$ and $\beta$ of the torus,
\begin{equation}
 \label{eq:periods}
 a = \oint_{\alpha} \frac{x\, dx}{y}, \qquad
    a_D = \oint_{\beta} \frac{x\, dx}{y}.
\end{equation}
The period $a$ serves as the local coordinate on the Coulomb branch $\mathcal{C}$,
while the dual period $a_D$ encodes the central charge of BPS states. The mass of
a dyon carrying electric charge $n_e$ and magnetic charge $n_m$ saturates the
BPS bound $M = |n_e \cdot a + n_m \cdot a_D|$.

The two periods are not independent: they are both derivatives of a single
holomorphic function --- the prepotential $\mathcal{F}(a)$ --- through the
relation $a_D = \partial\mathcal{F}/\partial a$. The complexified effective gauge
coupling then follows as
\begin{equation}
    \tau(u) = \frac{\partial a_D}{\partial a} = \frac{\partial^2 \mathcal{F}}{\partial a^2},
\end{equation}
where $\mathrm{Im}\,\tau(u) > 0$ is guaranteed by the positive-definiteness of the
low-energy kinetic term. 

The prepotential $\mathcal{F}(a)$ can be derived by three complementary methods,
each illuminating a different aspect of the theory. The first is a direct
perturbative calculation: owing to the non-renormalization theorems of
$\mathcal{N}=2$ supersymmetry, the perturbative series for $\mathcal{F}$ truncates
at one loop, yielding $\mathcal{F}_{\mathrm{1-loop}} \sim (i/2\pi)\,a^2 \ln(a^2/\Lambda^2)$,
where the logarithm arises from the standard one-loop momentum integral and
$\Lambda$ is the dynamically generated scale of the theory. The second method
exploits the $R$-symmetry and instanton-number selection rules: since $\mathcal{F}$
must have mass dimension two and the $k$-instanton sector of $\mathrm{SU}(2)$
contributes at order $\Lambda^{4k}$, dimensional analysis uniquely fixes the
structure of each term in the non-perturbative series to be proportional to
$a^2(\Lambda/a)^{4k}$, leading directly to the expansion
\begin{equation}
\label{eq:poten}
    \mathcal{F}(a) = \frac{i}{2\pi}\,a^2 \ln\frac{a^2}{\Lambda^2}
    + \sum_{k=1}^{\infty} \mathcal{F}_k
      \left(\frac{\Lambda}{a}\right)^{4k} a^2.
\end{equation}

The most direct and rigorous derivation, however, proceeds through the
\textbf{period integrals} of the Seiberg--Witten curve. From the definition
$a_D = \partial\mathcal{F}/\partial a$ and the explicit period formulae see~\eqref{eq:periods}
one can express $a_D$ as a function of $a$ and subsequently reconstruct
$\mathcal{F}(a)$ by integration. Expanding the curve in the weak-coupling regime
$a \gg \Lambda$ and evaluating the period integrals order by order in $\Lambda/a$
reproduces precisely the same formula, with the instanton coefficients
$\mathcal{F}_k$ now determined \textbf{exactly} --- without any approximation or
truncation. The complete agreement between all three approaches constitutes a
highly non-trivial consistency check and underscores the internal coherence of the
Seiberg--Witten solution.

The geometry of $\mathcal{C}$ is that of a \textbf{Special K\"{a}hler manifold},
defined by a flat $\mathrm{Sp}(2,\mathbb{Z})$ bundle whose holomorphic sections
encode the special coordinates $(a_D, a)$ and the holomorphic gauge couplings
$\tau_{ij} = \partial (a_D)_i/\partial a^j$. The non-trivial monodromy of this bundle
around metric singularities of $\mathcal{C}$ reflects the discrete electromagnetic
duality of the low-energy $\mathrm{U}(1)$ theory and has direct physical
consequences: at the singular loci, additional states become massless, invalidating
the effective field theory description.

\subsection{BPS states}
The electric and magnetic charges arise from independent sources. The electric
charge $n_e$ is the standard $\mathrm{U}(1)$ charge of the residual Abelian
gauge theory. The magnetic charge $n_m$, by contrast, is of topological origin:
magnetic monopoles appear as classical solitonic solutions with non-trivial
field topology,
\begin{equation}
    \pi_2\!\left(\mathrm{SU}(2)/\mathrm{U}(1)\right) =
    \pi_2(S^2) = \mathbb{Z},
\end{equation}
and $n_m$ is the associated Pontryagin number, i.e.\ the winding number of the
field configuration at spatial infinity. Together, the charges form the
\textbf{Dirac--Schwinger--Zwanziger lattice} $\Gamma = \mathbb{Z}^2$, equipped
with the symplectic pairing
\begin{equation}
    \langle (n_e, n_m),\, (n_e', n_m') \rangle
    = n_e n_m' - n_m n_e',
\end{equation}
which must be an integer by the Dirac quantization condition. The existence of
this lattice structure is the algebraic origin of the $\mathrm{SL}(2,\mathbb{Z})$
duality group, since $\mathrm{SL}(2,\mathbb{Z})$ is precisely the group of
automorphisms of $\mathbb{Z}^2$ that preserve the symplectic form.

In $\mathcal{N}=2$ supersymmetry algebra, the anticommutator of supercharges
contains a central charge $Z = n_e \cdot a + n_m \cdot a_D$, which directly
enters the BPS bound above. States saturating this bound are absolutely stable
against decay at fixed charges, and it is these BPS states whose masses are
computed exactly by the period integrals of $\lambda_{\mathrm{SW}}$.
The spectrum of BPS states in $SU(2)$ $\mathcal{N}=2$ Seiberg--Witten theory consists of electrically charged particles, monopoles, and dyons. Their masses satisfy the BPS bound \cite{WittenOlive,HarveyMoore}

\begin{equation} M = |Z| = |n_e a + n_m a_D|. \end{equation}

The basic charges can be summarized in Table~\ref{tab:bps}.

\begin{table}[ht]
\centering
\caption{Basic BPS states in $SU(2)$ $\mathcal{N}=2$ gauge theory.}
\begin{tabular}{ccc}
\hline
State & $(n_e,n_m)$ & Description \\
\hline
Electron & $(1,0)$ & Fundamental electric excitation \\
Monopole & $(0,1)$ & Magnetic monopole \\
Dyon & $(1,1)$ & Dyonic BPS particle \\
Dyon & $(1,-1)$ & Dual dyon state \\
\hline
\end{tabular}
\label{tab:bps}
\end{table}

The stability of these states depends on the position in the $u$-plane. Near the monopole singularity the light state has charge $(0,1)$, while near the dyon point the light state carries charge $(1,-1)$.

\subsection{Picard–Fuchs System for Seiberg–Witten Periods}
The nontrivial structure of the solutions to the Picard--Fuchs equation \cite{MorrisonPF}
\[
\frac{d^2 a(u)}{du^2} + \frac{1}{4\left(u^2 - \Lambda^4\right)}\, a(u) = 0
\]
is determined by its behavior near the regular singular points. This is a second-order equation $\Rightarrow$ the solution space is two-dimensional. We choose a basis:
\[
\vec{\Pi}(u) = \begin{pmatrix} a(u) \\ a_D(u) \end{pmatrix}.
\]

\textbf{Monodromy} is how this vector transforms upon encircling a singular point \cite{Seiberg1994b}:
\[
\vec{\Pi} \longrightarrow M \vec{\Pi}, \quad M \in SL(2, \mathbb{Z}).
\]
In a neighborhood of \(u_0=\Lambda^2\), introducing a local coordinate \(\epsilon = u - \Lambda^2\), the equation reduces to a Fuchsian form with a simple pole. Applying the Frobenius method, one finds two independent solutions with indices \(\rho_1=0\) and \(\rho_2=1\), leading to the local behavior
\[
a_1(\epsilon) \sim 1, \qquad a_2(\epsilon) \sim \epsilon \log \epsilon.
\]

The appearance of the logarithmic term signals a nontrivial monodromy. Under analytic continuation around the singular point, \(\epsilon \to e^{2\pi i}\epsilon\), the logarithm transforms as
\[
\log \epsilon \to \log \epsilon + 2\pi i,
\]
which implies
\[
a_2 \to a_2 + 2\pi i\, a_1, \qquad a_1 \to a_1.
\]

Thus, in the basis of solutions \((a_D, a)\), where the logarithmic solution is identified with \(a_D\), the monodromy acts as
\[
\begin{pmatrix}
a_D \\
a
\end{pmatrix}
\;\longrightarrow\;
\begin{pmatrix}
1 & 1 \\
0 & 1
\end{pmatrix}
\begin{pmatrix}
a_D \\
a
\end{pmatrix},
\]
after an appropriate normalization. This construction generalizes to other singular points, where the corresponding local analysis determines the associated \(SL(2,\mathbb{Z})\) monodromy matrices.

\begin{figure}[htbp]
    \centering
    \includegraphics[width=0.75\textwidth]{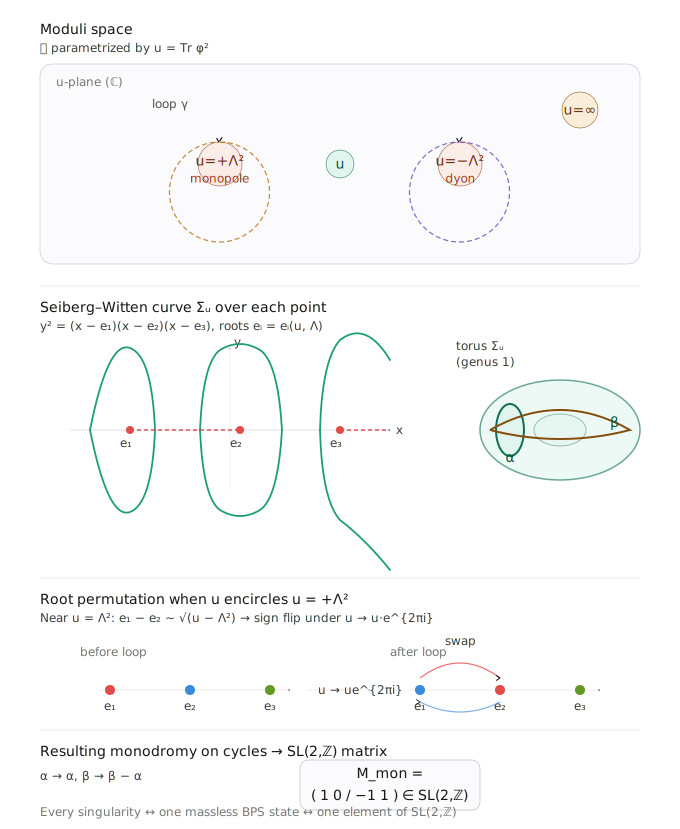}
    \caption[Geometric origin of monodromy in Seiberg--Witten theory]{%
    Geometric origin of monodromy in Seiberg--Witten theory.
    The moduli space $\mathcal{C}$ contains singular points at
    $u=\pm\Lambda^2$ and $u=\infty$, where BPS states become massless.
    A loop $\gamma$ encircling $u=\Lambda^2$ induces a non-trivial
    monodromy on the period vector $(a_D,a)$.
    
    \par\smallskip
    
   - The Seiberg--Witten curve $\Sigma_u$ is an elliptic curve
    $y^2=(x-e_1)(x-e_2)(x-e_3)$ whose homology cycles $(\alpha,\beta)$
    define the periods
    $a=\oint_\alpha \lambda_{\mathrm{SW}}$ and
    $a_D=\oint_\beta \lambda_{\mathrm{SW}}$.
    
    \par\smallskip
    
    - Near $u=\Lambda^2$ two branch points collide,
    $e_1-e_2\sim\sqrt{u-\Lambda^2}$, and are exchanged under analytic
    continuation $u\to u e^{2\pi i}$. This induces a transformation of
    the homology basis $(\alpha,\beta)$ and hence of the period vector.
    
    \par\smallskip
    
    - The resulting monodromy is
    $(a_D,a)\to(a_D,a)M$, with
    $M=\begin{pmatrix}1&0\\-1&1\end{pmatrix}\in SL(2,\mathbb{Z})$.%
    }
    \label{fig:SW_monodromy}
\end{figure}

As $u$ varies along a non-trivial loop in $\mathcal{M}$, the cycles $\alpha$ and
$\beta$ on the torus undergo a monodromy transformation, causing the periods to
transform.
The monodromy group of the periods coincides with the electromagnetic duality
group $\mathrm{SL}(2,\mathbb{Z})$. The geometric origin of electromagnetic duality can be understood in terms of the
monodromy of the Seiberg--Witten periods over the moduli space.
As one moves around singular loci in the $u$-plane, the homology cycles of the
elliptic curve undergo non-trivial transformations, which induce an action on
the period vector $(a_D,a)$. This structure is illustrated schematically in figure~\ref{fig:SW_monodromy}.

Every singularity of the moduli space
thus corresponds to a specific element of $\mathrm{SL}(2,\mathbb{Z})$,
determined by the charges of the massless BPS state at that locus.

\subsection{The Coulomb Branch and Its Geometry}

For $\Nequals{2}$ $\mathrm{SU}(2)$ Yang--Mills, the gauge group is broken
to $\mathrm{U}(1)$ by $\myvev{\phi}\propto\mathrm{diag}(a,-a)$.
The quantum Coulomb branch $\CB$ is a one-dimensional complex manifold
with a globally well-defined coordinate
\begin{equation}
  u = \mathrm{Tr}\,\phi^2 = 2a^2,
\end{equation}
whereas $a = \sqrt{u/2}$ is only \emph{locally} defined: it is a section
of a line bundle over $\CB$, not a function on it.  Under the monodromy
$\gamma$ around $u=0$,
\begin{equation}
  a \xrightarrow{\;\gamma\;} -a
  \quad\text{(residual Weyl group action $\mathbb{Z}_2$)}.
\end{equation}

The $\Nequals{2}$ effective action is completely determined by a single
holomorphic function — the \textbf{prepotential} $\MF(a)$:
\begin{equation}
  a_D = \frac{\partial\MF}{\partial a},
  \qquad
  \tau(a) = \frac{\partial^2\MF}{\partial a^2}
           = \frac{\theta_{\mathrm{YM}}}{2\pi}
           + \frac{4\pi i}{g^2_{\mathrm{eff}}(a)}.
\end{equation}
The metric on $\CB$ is \emph{special K\"{a}hler} \cite{StromingerSK}:
\begin{equation}
  ds^2 = \mathrm{Im}[\tau(a)]\,da\,d\bar{a},
  \label{eq:CBmetric}
\end{equation}
and the K\"{a}hler potential is $K = \mathrm{Im}(\bar a\, a_D)$.
\subsection{Confinement via Soft Breaking to \texorpdfstring{$\Nequals{1}$}{\textit{N}=1}}

Adding a tree-level mass $\mu\,\mathrm{Tr}\,\phi^2$ softly breaks
$\Nequals{2}\to\Nequals{1}$.  The theory dynamically selects vacua near
$u=\pm\Lambda^2$, where a monopole or dyon BPS state becomes massless and
condenses:
\begin{equation}
  \myvev{M}\neq 0 \quad\Longrightarrow\quad
  \text{dual Higgs mechanism}
  \quad\Longrightarrow\quad
  \text{colour confinement (electric flux tubes)}.
\end{equation}
This is the first \emph{microscopic} and exact proof of confinement in
four-dimensional gauge theory.

The entire low-energy dynamics is encoded geometrically:

\begin{equation}
\boxed{
\begin{aligned}
\text{vacua} &\leftrightarrow \text{base } \mathcal{M}, \\
\text{fields} &\leftrightarrow \lambda_{\mathrm{SW}}, \\
\text{charges} &\leftrightarrow H_1(\Sigma_u), \\
\text{masses} &\leftrightarrow \text{periods}, \\
\text{duality} &\leftrightarrow \text{monodromy}, \\
\text{strong coupling} &\leftrightarrow \text{degeneration of } \Sigma_u.
\end{aligned}
}
\end{equation}

Thus, the problem of determining the exact non-perturbative dynamics reduces to the study of a family of algebraic curves and their period integrals.

%%%%%%%%%%%%%%%%%%%%%%%%%%%%%%%%%%%%%%%%%%%%%%%%%%%%
\section{From Global Supersymmetry to Supergravity}\label{sec:SUGRA}

\subsection{Components of Global SUSY}
The compact superspace action~\eqref{eq:action_global} is not merely a
convenient shorthand: it is a \emph{generating functional} from which every
sector of the $\mathcal{N}=1$ Lagrangian is derived by a systematic and
unambiguous procedure. Once $K$ and $W$ are specified, no further choices
remain---the kinetic terms, Yukawa couplings, scalar potential, and gauge
interactions are all fixed. We now make this derivation explicit after expanding each superfield in powers of the Grassmann coordinates.

A chiral superfield $\Phi^{i}$, satisfying the constraint
$\bar{D}_{\dot{\alpha}}\Phi^{i}=0$, admits the component expansion
\begin{equation}
  \Phi^{i}(x,\theta,\bar{\theta})
  = \phi^{i}(x) + \sqrt{2}\,\theta^{\alpha}\psi^{i}_{\alpha}(x)
    + \theta^{2}F^{i}(x) + \ldots,
  \label{eq:chiral_expansion}
\end{equation}
where $\phi^{i}$ is a complex scalar, $\psi^{i}_{\alpha}$ is a left-handed
Weyl fermion, and $F^{i}$ is a complex auxiliary scalar with no propagating
degrees of freedom. The vector superfield $V^{a}$, restricted to the
Wess--Zumino gauge, expands as
\begin{equation}
  V^{a}(x,\theta,\bar{\theta})
  = -\theta\sigma^{\mu}\bar{\theta}\,A^{a}_{\mu}
    + i\theta^{2}\bar{\theta}\bar{\lambda}^{a}
    - i\bar{\theta}^{2}\theta\lambda^{a}
    + \tfrac{1}{2}\theta^{2}\bar{\theta}^{2}D^{a},
  \label{eq:vector_expansion}
\end{equation}
where $A^{a}_{\mu}$ is the gauge potential, $\lambda^{a}$ is the gaugino, and
$D^{a}$ is a real auxiliary scalar. The Grassmann integration rules
\begin{equation}
  \int d^{2}\theta\;(\cdots)
  \;\longrightarrow\;
  \bigl[\theta^{2}\bigr]\text{-component},
  \qquad
  \int d^{4}\theta\;(\cdots)
  \;\longrightarrow\;
  \bigl[\theta^{2}\bar{\theta}^{2}\bigr]\text{-component}
  \label{eq:Berezin}
\end{equation}
then extract the relevant piece of each superspace integrand. We now perform the Grassmann integral, which by the rules of Berezin integration simply selects a particular component of the resulting expansion in the superspace action

\begin{equation}
  S = \int d^{4}x\,d^{4}\theta\;K\!\left(\Phi^{i},\Phi^{\dagger}_{i}e^{2gV}\right)
    + \int d^{4}x\,d^{2}\theta\;W(\Phi^{i}) + \mathrm{h.c.}
    + \frac{1}{4g^{2}}\int d^{4}x\,d^{2}\theta\;
      \mathrm{Tr}(W^{\alpha}W_{\alpha}) + \mathrm{h.c.}
  \label{eq:master_action}
\end{equation}

% --------------------------------------------------------------------

{\bf{The kinetic sector}} originates entirely from the first term of~\eqref{eq:master_action}.
Substituting the expansion~\eqref{eq:chiral_expansion} into $K(\Phi^{i},\Phi^{\dagger}_{i})$
and expanding in a Taylor series around the scalar field values $\phi^{i}$,
\begin{equation}
  K(\Phi^{i},\Phi^{\dagger}_{i})
  = K(\phi^{i},\bar{\phi}^{\bar{\jmath}})
    + \partial_{i}K\,\delta\phi^{i}
    + \partial_{\bar{\jmath}}K\,\delta\bar{\phi}^{\bar{\jmath}}
    + \partial_{i}\partial_{\bar{\jmath}}K\,
      \delta\phi^{i}\delta\bar{\phi}^{\bar{\jmath}}
    + \ldots,
  \label{eq:K_Taylor}
\end{equation}
the $\theta^{2}\bar{\theta}^{2}$ coefficient of $K(\Phi,\Phi^{\dagger}e^{2gV})$,
selected by $\int d^{4}\theta$, yields
\begin{equation}
  \int d^{4}\theta\; K
  \;\longrightarrow\;
  g_{i\bar{\jmath}}\Bigl[
    \partial_{\mu}\phi^{i}\,\partial^{\mu}\bar{\phi}^{\bar{\jmath}}
    + i\,\bar{\psi}^{\bar{\jmath}}\bar{\sigma}^{\mu}D_{\mu}\psi^{i}
    + F^{i}\bar{F}^{\bar{\jmath}}
  \Bigr],
  \label{eq:sector1_raw}
\end{equation}
where the K\"{a}hler metric $g_{i\bar{\jmath}}=\partial_{i}\partial_{\bar{\jmath}}K$
arises as the coefficient of $\delta\phi^{i}\delta\bar{\phi}^{\bar{\jmath}}$ in
the Taylor expansion~\eqref{eq:K_Taylor}. The term $F^{i}\bar{F}^{\bar{\jmath}}$
in~\eqref{eq:sector1_raw} is auxiliary and will be eliminated below.

% --------------------------------------------------------------------

% --------------------------------------------------------------------

{\bf{The gauge kinetic sector}} originates from the third term of~\eqref{eq:master_action}.
The gauge-covariant field-strength superfield is defined as
$W_{\alpha}^{a}=-\tfrac{1}{4}\bar{D}^{2}(e^{-2gV}D_{\alpha}e^{2gV})^{a}$,
and its square evaluates in components as
\begin{equation}
  \mathrm{Tr}(W^{\alpha}W_{\alpha})\big|_{\theta^{2}}
  = -\tfrac{1}{2}F^{a}_{\mu\nu}F^{a\mu\nu}
    + \tfrac{i}{2}F^{a}_{\mu\nu}\tilde{F}^{a\mu\nu}
    + 2i\bar{\lambda}^{a}\bar{\sigma}^{\mu}D_{\mu}\lambda^{a}
    + D^{a}D^{a},
  \label{eq:WW_component}
\end{equation}
where $\tilde{F}^{a\mu\nu}=\tfrac{1}{2}\varepsilon^{\mu\nu\rho\sigma}F^{a}_{\rho\sigma}$.
In the general case, the gauge kinetic term $\tfrac{1}{4g^{2}}\to\tfrac{1}{4}f_{ab}(\phi)$
introduces the holomorphic gauge kinetic function $f_{ab}(\phi)$, which depends
on the chiral scalars and therefore varies over the moduli space. This gives
\begin{equation}
  \mathcal{L}_{\mathrm{gauge}}
  = \tfrac{1}{4}\,\mathrm{Re}(f_{ab})\,F^{a}_{\mu\nu}F^{b\mu\nu}
    + \tfrac{1}{4}\,\mathrm{Im}(f_{ab})\,F^{a}_{\mu\nu}\tilde{F}^{b\mu\nu}
    + i\,\mathrm{Re}(f_{ab})\,\bar{\lambda}^{a}\bar{\sigma}^{\mu}D_{\mu}\lambda^{b}.
  \label{eq:sector2}
\end{equation}
The real part of $f_{ab}$ sets the effective gauge coupling
$g_{\mathrm{eff}}^{-2}=\mathrm{Re}(f_{ab})$, while its imaginary part
multiplies the topological density and determines the $\theta$-angle. Both
depend on the vacuum expectation values of $\phi^{i}$, so the geometry of the
scalar manifold $\mathcal{M}$ directly controls the gauge dynamics.

% --------------------------------------------------------------------

% --------------------------------------------------------------------

{\bf{The auxiliary fields}} $F^{i}$ and $D^{a}$ carry no kinetic terms and therefore
satisfy purely algebraic equations of motion. These equations link the
auxiliary fields to derivatives of $K$ and $W$, and substituting the solutions
back into the Lagrangian generates the scalar potential.

{\it{F-term potential}}

Collecting all terms involving $F^{i}$ from~\eqref{eq:sector1_raw} and from
the superpotential contribution $\int d^{2}\theta\,W\big|_{\theta^{2}}=\partial_{i}W\,F^{i}$,
the Lagrangian contains
\begin{equation}
  \mathcal{L} \supset g_{i\bar{\jmath}}\,F^{i}\bar{F}^{\bar{\jmath}}
  + \partial_{i}W\,F^{i} + \overline{\partial_{i}W}\,\bar{F}^{\bar{\imath}}.
  \label{eq:aux_F_terms}
\end{equation}
Varying with respect to $\bar{F}^{\bar{\imath}}$ gives
$g_{i\bar{\jmath}}\,F^{i}=-\overline{\partial_{\bar{\jmath}}W}$, hence
\begin{equation}
  F^{i} = -g^{i\bar{\jmath}}\,\overline{\partial_{\bar{\jmath}}W}.
  \label{eq:F_eom}
\end{equation}
Substituting~\eqref{eq:F_eom} into~\eqref{eq:aux_F_terms} and changing sign
to obtain the potential,
\begin{equation}
  V_{F} = g^{i\bar{\jmath}}\,\partial_{i}W\,\overline{\partial_{\bar{\jmath}}W}
         = \sum_{i}\left|\frac{\partial W}{\partial\phi^{i}}\right|^{2}
         \;\geq\; 0.
  \label{eq:VF_global}
\end{equation}
This is manifestly non-negative in the global theory, a property that will be
lost upon coupling to gravity.

{\it{D-term potential}}

From~\eqref{eq:WW_component} and the matter--gauge coupling generated by
$\int d^{4}\theta\,K(\Phi^{\dagger}e^{2gV},\Phi)$, the auxiliary D-field
appears as
\begin{equation}
  \mathcal{L} \supset \tfrac{1}{2}\,D^{a}D^{a}
  + g\,D^{a}\,\partial_{i}K\,(T_{a})^{i}{}_{j}\phi^{j}.
  \label{eq:aux_D_terms}
\end{equation}
The algebraic equation of motion $D^{a}=-g\,\partial_{i}K\,(T_{a})^{i}{}_{j}\phi^{j}$
substituted back yields
\begin{equation}
  V_{D} = \tfrac{1}{2}(\mathrm{Re}\,f)^{-1\,ab}\,D_{a}D_{b},
  \qquad
  D_{a} = g\,\partial_{i}K\,(T_{a})^{i}{}_{j}\phi^{j},
  \label{eq:VD}
\end{equation}
so the total scalar potential in the global theory is
$V=V_{F}+V_{D}\geq 0$.

% --------------------------------------------------------------------

% --------------------------------------------------------------------

{\bf{The Yukawa sector}} arises from two independent sources in~\eqref{eq:master_action},
and it is instructive to trace each contribution separately.

{\it{Superpotential Yukawa couplings}}

Expanding $W(\Phi^{i})$ in~\eqref{eq:chiral_expansion} to quadratic order in
$\theta$,
\begin{equation}
  W(\Phi^{i}) = W(\phi^{i})
    + \frac{\partial W}{\partial\phi^{i}}\sqrt{2}\,\theta\psi^{i}
    + \frac{\partial^{2}W}{\partial\phi^{i}\partial\phi^{j}}
      \theta\psi^{i}\,\theta\psi^{j} + \ldots
  \label{eq:W_expansion}
\end{equation}
and applying $\int d^{2}\theta$, which selects the $\theta^{2}$ component,
\begin{equation}
  \int d^{2}\theta\;W(\Phi^{i})
  \;\longrightarrow\;
  \partial_{i}W\cdot F^{i}
  - \tfrac{1}{2}\,\partial_{i}\partial_{j}W\;\psi^{i}\psi^{j}.
  \label{eq:W_component}
\end{equation}
The first term was used above to derive the $F$-term potential~\eqref{eq:VF_global}. The second term gives the holomorphic fermion bilinear
\begin{equation}
  \mathcal{L}_{\mathrm{Yuk},\,W}
  = -\tfrac{1}{2}\,\partial_{i}\partial_{j}W\;\psi^{i}\psi^{j} + \mathrm{h.c.},
  \label{eq:YukawaW}
\end{equation}
which generates fermion masses and Yukawa vertices once $W$ is specified.

{\it{Gauge Yukawa couplings}}

Two independent sources contribute to the gauge Yukawa sector.
The first arises from the K\"{a}hler potential
$K(\Phi^{\dagger}e^{2gV},\Phi)$ upon $\int d^{4}\theta$, yielding
the standard matter--gaugino coupling:
\begin{equation}\label{eq:YukawaK}
  \mathcal{L}_{\mathrm{Yuk},\,K}
  = -\sqrt{2}\,g\,\phi^{\dagger}_{i}(T^{a})^{i}{}_{j}\,\psi^{j}\lambda^{a}
    + \mathrm{h.c.}
\end{equation}
The second arises from the holomorphic gauge kinetic function
$f_{ab}(\Phi)$ upon $\int d^{2}\theta\,f_{ab}W^{a\alpha}W^{b}_\alpha$,
generating an interaction between matter fermions and two gauginos:
\begin{equation}\label{eq:YukawaGauge}
  \mathcal{L}_{\mathrm{Yuk},\,f}
  = -\tfrac{1}{\sqrt{2}}\,(\partial_{i}f_{ab})\,\psi^{i}\lambda^{a}\lambda^{b}
    + \mathrm{h.c.}
\end{equation}

Therefore, Yukawa sector receives contributions from three
distinct sources and can be written as
\begin{equation}\label{eq:sector4}
  \mathcal{L}_{\mathrm{Yukawa}}
  = \underbrace{-\tfrac{1}{2}\,\partial_{i}\partial_{j}W\,\psi^{i}\psi^{j}
    }_{\text{from }W(\Phi)}
  \underbrace{-\sqrt{2}\,g\,\phi^{\dagger}_{i}
    (T^{a})^{i}{}_{j}\,\psi^{j}\lambda^{a}
    }_{\text{from }K(\Phi^{\dagger}e^{2gV},\Phi)}
  \underbrace{-\tfrac{1}{\sqrt{2}}\,(\partial_{i}f_{ab})\,
    \psi^{i}\lambda^{a}\lambda^{b}
    }_{\text{from }f_{ab}(\Phi)}
  \;+\;\mathrm{h.c.}
\end{equation}
The first term generates fermion masses and trilinear Yukawa vertices
entirely from the holomorphic superpotential $W$, and is present even
in the ungauged theory. The second term, proportional to the gauge
coupling $g$, describes the interaction between a matter scalar
$\phi^\dagger_i$, a matter fermion $\psi^j$, and a gaugino $\lambda^a$;
it arises from the covariant coupling $K(\Phi^{\dagger}e^{2gV},\Phi)$
upon performing the $\int d^{4}\theta$ integral. The third term couples
a matter fermion $\psi^i$ to two gauginos $\lambda^a\lambda^b$ with
a coefficient given by the derivative of the holomorphic gauge kinetic
function $f_{ab}(\Phi)$ with respect to the chiral scalars; it vanishes
for a constant $f_{ab} = \delta_{ab}/g^2$ but becomes non-trivial in
models where the gauge kinetic function depends on moduli fields, as
is the case in supergravity and string compactifications. Together,
these three terms account for all renormalisable Yukawa interactions
permitted by $\mathcal{N}=1$ supersymmetry.

% --------------------------------------------------------------------

% --------------------------------------------------------------------

{\bf{The fifth and final sector}} has no counterpart in non-supersymmetric theories
and is often overlooked in elementary treatments. When $\int d^{4}\theta\,K$
is expanded to quartic order in the fermionic components $\psi^{i}$, a contact
four-fermion interaction is generated whose coefficient is the Riemann
curvature tensor of the scalar manifold $\mathcal{M}$:
\begin{equation}
  \mathcal{L}_{4f}
  = -R_{i\bar{\jmath}k\bar{l}}\;
    \psi^{i}\psi^{k}\bar{\psi}^{\bar{\jmath}}\bar{\psi}^{\bar{l}}.
  \label{eq:sector5}
\end{equation}
Here $R_{i\bar{\jmath}k\bar{l}}=\partial_{i}\partial_{\bar{\jmath}}g_{k\bar{l}}
-g^{m\bar{n}}\Gamma_{ik}^{m}\overline{\Gamma_{\bar{\jmath}\bar{l}}^{n}}$
is the Riemann tensor of $\mathcal{M}$ computed from the K\"{a}hler metric
\begin{equation}
g_{i\bar{j}} = \frac{\partial^2 K}{\partial \phi^i \partial \bar{\phi}^{\bar{j}}} \equiv \partial_i \partial_{\bar{j}} K.
\label{eq:kahler_metric}
\end{equation}
This is a Hermitian metric on the complex manifold of scalar fields. The associated Levi-Civita connection (K\"ahler connection) on the holomorphic tangent bundle is:

\begin{equation}
\Gamma^k_{ij} = g^{k\bar{l}} \partial_i g_{j\bar{l}} = g^{k\bar{l}} \partial_i \partial_j \partial_{\bar{l}} K.
\end{equation}
It is therefore absent in global
$\mathcal{N}=1$ supersymmetry with canonical kinetic terms, but becomes mandatory in
supergravity and in non-linear sigma models defined on curved K\"{a}hler target spaces.
In practice, this term contributes to four-scalar scattering amplitudes at tree level
and provides a direct geometric probe of the curvature of the scalar field space.
% --------------------------------------------------------------------

The derivation above shows that the master action~\eqref{eq:master_action}
is a genuine generating functional: specifying the three holomorphic functions
$K$, $W$, and $f_{ab}$ completely and uniquely determines the entire
Lagrangian. No further freedom remains. Table~\ref{tab:derivation} organises
the five sectors by their origin within~\eqref{eq:master_action} and the
operation that extracts them.

\begin{table}[ht]
\centering
\caption{Derivation of the five Lagrangian sectors from the master
         action~\eqref{eq:master_action}.}
\label{tab:derivation}
\renewcommand{\arraystretch}{1.5}
\begin{tabular}{llll}
\toprule
\textbf{Sector} & \textbf{Source in~\eqref{eq:master_action}}
                & \textbf{Operation} & \textbf{Result} \\
\midrule
Kinetic ($\mathcal{L}_{\mathrm{kin}}$)
  & $\int d^{4}\theta\,K$
  & $[\theta^{2}\bar{\theta}^{2}]$, quadratic in fields
  & Eq.~\eqref{eq:sector1_raw} \\
Gauge ($\mathcal{L}_{\mathrm{gauge}}$)
  & $\int d^{2}\theta\,\mathrm{Tr}(W^{\alpha}W_{\alpha})$
  & $[\theta^{2}]$ of $W^{\alpha}W_{\alpha}$
  & Eq.~\eqref{eq:sector2} \\
Potential ($V_{F}+V_{D}$)
  & both $K$ and $W$ terms
  & eliminate $F^{i}$, $D^{a}$ on-shell
  & Eqs.~\eqref{eq:VF_global},\,\eqref{eq:VD} \\
Yukawa ($\mathcal{L}_{\mathrm{Yukawa}}$)
  & $\int d^{2}\theta\,W$ and $\int d^{4}\theta\,K$
  & $[\theta^{2}]$ of $W$; $[\theta^{2}\bar{\theta}^{2}]$ of $K$
  & Eq.~\eqref{eq:sector4} \\
Four-fermion ($\mathcal{L}_{4f}$)
  & $\int d^{4}\theta\,K$
  & $[\theta^{2}\bar{\theta}^{2}]$, quartic in fermions
  & Eq.~\eqref{eq:sector5} \\
\bottomrule
\end{tabular}
\end{table}

% ====================================================================
\subsection{The Transition to Supergravity: Local Geometry}
\label{sec:local_geometry}
% ====================================================================

Having established that the master action~\eqref{eq:master_action} generates
all sectors of the globally supersymmetric Lagrangian, we now analyse what
changes when supersymmetry is promoted from a global to a local symmetry.
The transition is conceptually clean and proceeds through three logically
distinct steps.

% --------------------------------------------------------------------
{\bf{Step I: Localisation of the Supersymmetry Parameter}}
\label{subsec:step1}
% --------------------------------------------------------------------

In the global theory the supersymmetry parameter $\epsilon_{\alpha}$ is a
constant spinor: it is the same at every spacetime point. When one attempts
to make this parameter position-dependent,
\begin{equation}
  \epsilon_{\alpha} = \mathrm{const}
  \;\longrightarrow\;
  \epsilon_{\alpha}(x),
  \label{eq:localisation_param}
\end{equation}
the ordinary partial derivatives $\partial_{\mu}$ appearing in the
transformation rules of all fields cease to be covariant. This is precisely
the same obstruction encountered in electrodynamics when a global $U(1)$
phase is made local: the resolution there was to introduce a gauge potential
$A_{\mu}$ and replace $\partial_{\mu}\to\partial_{\mu}-ieA_{\mu}$. By the
same logic, consistency of the locally supersymmetric transformation rules
requires the introduction of a gauge field for the fermionic symmetry. This
new field is the \emph{gravitino} $\psi_{\mu\alpha}$, a spin-$\tfrac{3}{2}$
Rarita--Schwinger field. Since the supersymmetry algebra closes on spacetime
translations, the gauge field of the fermionic symmetry necessarily brings
with it the gauge field of spacetime translations---the \emph{graviton}
$g_{\mu\nu}$ (spin~$2$). The two together form the supergravity multiplet:
\begin{equation}
  \text{global SUSY},\;\epsilon_{\alpha}=\mathrm{const}
  \;\xrightarrow{\;\epsilon_{\alpha}\to\epsilon_{\alpha}(x)\;}
  \;\mathcal{N}=1\;\text{supergravity},\quad
  \text{multiplet:}\;(g_{\mu\nu},\,\psi_{\mu\alpha}).
  \label{eq:SUGRA_multiplet}
\end{equation}
The vierbein $e^{a}{}_{\mu}$, defined through $g_{\mu\nu}=e^{a}{}_{\mu}e^{b}{}_{\nu}\eta_{ab}$,
encodes the metric, while the gravitino plays the role of the gauge potential
of local supersymmetry, in complete analogy with the role of $A_{\mu}$ in
gauge theory.

% --------------------------------------------------------------------
{\bf{Step II: Curved Superspace and the Deformation of the Measure}}
\label{subsec:step2}
% --------------------------------------------------------------------

The second and most profound change concerns the geometric structure of
superspace itself. In the global theory the superspace measure is flat:
$d^{4}x\,d^{4}\theta$, with $d^{4}x=dx^{0}dx^{1}dx^{2}dx^{3}$ being the
ordinary Lebesgue measure on Minkowski space. Upon localisation, spacetime
becomes curved, and the correct generally covariant measure is
$d^{4}x\sqrt{-g}$. In superspace this generalises to
\begin{equation}
  d^{4}x\,d^{4}\theta
  \;\longrightarrow\;
  d^{4}x\,d^{4}\theta\; E,
  \label{eq:measure_deformation}
\end{equation}
where $E=\mathrm{sdet}(E^{A}{}_{M})$ is the superdeterminant of the
supervielbein $E^{A}{}_{M}$---the supersymmetric generalisation of the
vierbein. Likewise, the chiral measure is deformed as
$d^{4}x\,d^{2}\theta\to d^{4}x\,d^{2}\theta\,\mathcal{E}$, where
$\mathcal{E}$ is the chiral projection of $E$. The master
action~\eqref{eq:master_action} thus becomes \cite{WessBagger1992,FreedmanVanProeyen}
\begin{equation}
  S_{\mathrm{SUGRA}}
  = \int d^{4}x\,d^{4}\theta\; E\cdot K\!\left(\Phi^{i},\Phi^{\dagger}_{i}e^{2gV}\right)
    + \int d^{4}x\,d^{2}\theta\;\mathcal{E}\cdot W(\Phi^{i})
    + \mathrm{h.c.}
  \label{eq:action_SUGRA}
\end{equation}
When the superdeterminant $E$ is expanded in component fields of the
supergravity multiplet, it introduces the metric determinant $\sqrt{-g}$
together with additional couplings to the spin connection and the gravitino.
Crucially, the expansion of $\mathcal{E}$ in components contains the
\emph{curvature superfield} $R$---the superspace scalar curvature---which
couples directly to the superpotential $W$. It is this coupling that produces
the negative term in the scalar potential, as we discuss in
Step~III below.

% --------------------------------------------------------------------
{\bf{Step III: Modification of K\"{a}hler Invariance and
            the Origin of the Negative Term}}
\label{subsec:step3}
% --------------------------------------------------------------------

In the global theory a K\"{a}hler transformation
leaves the action invariant without
any accompanying change in $W$. 
In the locally supersymmetric theory this is no longer the case: the curved
measure $E$ is not invariant under such a shift
\begin{equation}
K \to K + f(\phi) + \bar{f}(\bar{\phi}), \qquad f \text{ holomorphic},
\label{eq:kahler_transform}
\end{equation}
the metric $g_{i\bar{j}}$ is invariant because $\partial_i \partial_{\bar{j}} f = 0$. The SUGRA action contains the factor $e^{K/M_{\mathrm{Pl}}^2} |W|^2$, which is \textit{not} invariant under Eq.~\eqref{eq:kahler_transform} unless $W$ simultaneously transforms as:

\begin{equation}
W \to e^{-f(\phi)/M_{\mathrm{Pl}}^2} \, W.
\label{eq:W_transform}
\end{equation}
The unique K\"ahler-invariant combination of $K$ and $W$ is the \textit{K\"ahler function} $\mathcal{G}$:

\begin{equation}
\boxed{\mathcal{G}(\phi, \bar{\phi}) = \frac{K(\phi, \bar{\phi})}{M_{\mathrm{Pl}}^2} + \ln \frac{|W(\phi)|^2}{M_{\mathrm{Pl}}^6}}.
\label{eq:G_function}
\end{equation}
Indeed, under Eqs.~\eqref{eq:kahler_transform} and~\eqref{eq:W_transform}, the simultaneous shift of $\ln |W|^2$ cancels exactly, so $\mathcal{G} \to \mathcal{G}$.

\begin{remark}
$K$ and $W$ are individually unphysical (gauge artifacts of K\"ahler symmetry); only $\mathcal{G}$---or equivalently, K\"ahler-covariant derivatives of $W$---enters physical observables.
\end{remark}

{\bf{The SUGRA Scalar Potential}}
In terms of the scalar fields, the scalar potential takes the elegant form:

\begin{equation}
V = M_{\mathrm{Pl}}^4 \, e^{\mathcal{G}} \left( g^{i\bar{j}} \mathcal{G}_i \mathcal{G}_{\bar{j}} - 3 \right), \qquad \mathcal{G}_i \equiv \frac{\partial \mathcal{G}}{\partial \phi^i},
\label{eq:SUGRA_scalar_potential}
\end{equation}
where $g^{i\bar{j}}$ is the inverse K\"ahler metric. The K\"ahler-covariant derivative of $W$ is defined as:

\begin{equation}
D_i W = \partial_i W + \frac{1}{M_{\mathrm{Pl}}^2} (\partial_i K) W.
\end{equation}
Equivalently, the scalar potential can be written as:

\begin{equation}
V = e^{K/M_{\mathrm{Pl}}^2} \left( K^{i\bar{j}} D_i W D_{\bar{j}} \bar{W} - \frac{3}{M_{\mathrm{Pl}}^2} |W|^2 \right).
\label{eq:scalar_potential_alt}
\end{equation}
The exponential factor $e^{K/M_{\mathrm{Pl}}^2}$ and the negative term $-3|W|^2/M_{\mathrm{Pl}}^2$ are characteristic signatures of supergravity, absent in global SUSY. The transition to supergravity thus introduces two inseparable modifications
to the potential: a field-dependent overall rescaling and a gravitationally
induced negative contribution that allows the vacuum energy to take any sign.

As the result, the transition from global supersymmetry to supergravity involves several key modifications:

\begin{enumerate}
    \item \textbf{Spacetime geometry:} Flat Minkowski space $\eta_{\mu\nu}$ is replaced by curved spacetime $g_{\mu\nu}(x)$.
    
    \item \textbf{Superspace structure:} The superdeterminant $E = \text{sdet}(E_M^A)$ of the supervielbein replaces the flat superspace measure.
    
    \item \textbf{Planck scale:} The Planck mass $M_{\mathrm{Pl}}$ appears as the natural scale suppressing gravitational interactions.
    
    \item \textbf{K\"ahler invariance:} Physical observables depend only on the invariant combination $\mathcal{G}$, not on $K$ and $W$ separately.
    
    \item \textbf{Scalar potential:} The SUGRA potential \eqref{eq:scalar_potential} differs fundamentally from the global SUSY form $V = \sum_i |F_i|^2 + \frac{1}{2} \sum_a D_a^2$.
\end{enumerate}

% --------------------------------------------------------------------
\subsection{Modifications to All Five Sectors under Localisation}
\label{subsec:all_sectors_local}
% --------------------------------------------------------------------

The three steps above propagate through every sector of the Lagrangian,
replacing all global structures by their locally supersymmetric counterparts.
Table~\ref{tab:global_vs_local} records the precise modification in each
sector.

\begin{table}[ht]
\centering
\caption{Modification of each Lagrangian sector in the transition from global
         supersymmetry to supergravity.}
\label{tab:global_vs_local}
\renewcommand{\arraystretch}{1.6}
\begin{tabular}{lll}
\toprule
\textbf{Sector} & \textbf{Global SUSY} & \textbf{Supergravity} \\
\midrule
Spacetime measure
  & $d^{4}x$
  & $d^{4}x\sqrt{-g}$ \\
Superspace measure
  & $d^{4}x\,d^{4}\theta$
  & $d^{4}x\,d^{4}\theta\;E$ \\
Gravitational sector
  & absent
  & $\frac{1}{2\kappa^{2}}R
    +\bar{\psi}_{\mu}\gamma^{\mu\nu\rho}D_{\nu}\psi_{\rho}$ \\
Kinetic metric
  & $g_{i\bar{\jmath}}=\partial_{i}\partial_{\bar{\jmath}}K$
  & same; $\partial_{\mu}\to D_{\mu}$ (spin connection) \\
Yukawa coupling
  & $\partial_{i}\partial_{j}W$
  & $e^{K/2M_{\mathrm{Pl}}^{2}}\,D_{i}D_{j}W$
    (with $\Gamma^{k}{}_{ij}$) \\
Four-fermion term
  & $R_{i\bar{\jmath}k\bar{l}}\,\psi^{i}\psi^{k}
    \bar{\psi}^{\bar{\jmath}}\bar{\psi}^{\bar{l}}$
  & same + gravitino contributions \\
Scalar potential & $V = \sum |F_i|^2 + \frac{1}{2}\sum D_a^2$ & $V = \Mpl^4\,e^{\mathcal{G}}(g^{i\bar{j}}\mathcal{G}_i\mathcal{G}_{\bar{j}} - 3)$ \\
K\"{a}hler symmetry
  & $K\to K+f+\bar{f}$ trivial
  & requires $W\to e^{-f/M_{\mathrm{Pl}}^{2}}W$ \\
Physical data
  & $K$, $W$ separately
  & only $G=K/M_{\mathrm{Pl}}^{2}+\ln|W|^{2}/M_{\mathrm{Pl}}^{6}$ \\
Gravitino mass
  & $m_{3/2}=0$
  & $m_{3/2}=e^{K/2M_{\mathrm{Pl}}^{2}}|W|/M_{\mathrm{Pl}}^{2}$ \\
\bottomrule
\end{tabular}
\end{table}

The central lesson of Table~\ref{tab:global_vs_local} is that the geometry
of the scalar field manifold $\mathcal{M}$---defined in both theories by the
K\"{a}hler potential $K$---acquires a new role upon localisation: through the
curved superspace measure $E$ and the K\"{a}hler function $G$, the geometry
of $\mathcal{M}$ becomes coupled to the geometry of spacetime. The curvature
of $\mathcal{M}$ still enters the four-fermion interaction, and its geodesic
structure still organises the fermion mass matrices; but now these geometric
quantities are evaluated in a curved spacetime background, and the scalar
potential is no longer bounded below. The supergravity theory is therefore
not merely the global theory with gravity added: it is a genuinely new
framework in which two geometries---that of spacetime and that of the scalar
field manifold---are inextricably coupled through the holomorphic functions
$K$, $W$, and $f_{ab}$.

\subsection*{Conceptual Leap: From the Prepotential to K\"ahler Geometry}

In the $\mathcal{N}=2$ globally supersymmetric theory, the entire physical content of the 
Coulomb branch is encoded in a \emph{single} holomorphic function --- the prepotential 
$\mathcal{F}(a)$. The K\"ahler geometry of the moduli space is not an independent input 
but a \emph{consequence}: the metric, the effective coupling, and the symplectic structure 
of the charge lattice all follow from $\mathcal{F}$ by differentiation,
\begin{equation}
    a_D = \frac{\partial \mathcal{F}}{\partial a}, 
    \qquad
        \tau(u) = \frac{\partial^2 \mathcal{F}}{\partial a^2},
    \qquad
    ds^2=\mathrm{Im}[\tau(a)]dad{\overline{a}}
     \qquad
    K_{\mathrm{CulonBrunch}} = \mathrm{Im}\!\left(\bar{a}\, a_D\right).
\end{equation}
In $\mathcal{N}=1$ supergravity, the analogous role is played by \emph{two} holomorphic 
functions --- the K\"ahler potential $K(\Phi_i, \Phi^\dagger_i)$ and the superpotential 
$W(\Phi_i)$ --- which together determine all five sectors of the Lagrangian without any 
further freedom. The structural parallel between the two frameworks is summarised in 
Table~\ref{tab:n2_vs_sugra}.

\begin{table}[h]
\centering
\caption{Structural correspondence between $\mathcal{N}=2$ Seiberg--Witten theory 
and $\mathcal{N}=1$ supergravity. The prepotential $\mathcal{F}$ of the former 
generalises to the pair $(K, W)$ of the latter; special K\"ahler geometry of 
the Coulomb branch becomes the general K\"ahler geometry of the scalar manifold.}
\renewcommand{\arraystretch}{1.4}
\begin{tabular}{lll}
\hline
\textbf{Object} 
    & \textbf{$\mathcal{N}=2$ global SUSY} 
    & \textbf{$\mathcal{N}=1$ supergravity} \\
\hline
Generating function(s) 
    & prepotential $\mathcal{F}(a)$ 
    & K\"ahler potential $K$ and superpotential $W$ \\
Scalar geometry       
    & special K\"ahler (rigid) 
    & general K\"ahler manifold $\mathcal{M}$ \\
Vacuum parameter space 
    & Coulomb branch $\mathcal{C}$ 
    & scalar field manifold $\mathcal{M}$ \\
Metric tensor         
    & $\mathrm{Im}\,\tau(u)\,da\,d\bar{a}$ 
    & $g_{i\bar{\jmath}} = \partial_i \partial_{\bar{\jmath}} K$ \\
Physical coupling     
    & $\tau = \partial^2\mathcal{F}/\partial a^2$ 
    & $f_{ab}(\Phi)$ (holomorphic gauge kinetic function) \\
Scalar potential      
    & $V = 0$ on Coulomb branch 
    & $V = e^{K/M_{\mathrm{Pl}}^2}\!\left(K^{i\bar{\jmath}}D_i W\,\overline{D_j W} 
      - \tfrac{3}{M_{\mathrm{Pl}}^2}|W|^2\right)$ \\
\hline
\end{tabular}
\label{tab:n2_vs_sugra}
\end{table}

The deepest point of contact between the two frameworks lies in the role of 
\emph{holomorphy}. In the $\mathcal{N}=2$ theory, the non-renormalisation theorems 
guarantee that $\mathcal{F}$ receives contributions only at one loop in perturbation 
theory and from instanton sectors; its holomorphic dependence on $a$ is exact. 
In $\mathcal{N}=1$ supergravity, the superpotential $W(\Phi_i)$ is likewise protected 
by holomorphy: it is not renormalised in perturbation theory, and quantum corrections 
are restricted to non-perturbative contributions. Both structures therefore exemplify 
the same organising principle:
\begin{equation}
\boxed{%
    \text{vacuum structure of a supersymmetric theory} 
    \;\longleftrightarrow\; 
    \text{geometry of a holomorphic object.}
}
\end{equation}
The transition from Section~\ref{sec:SW} to Section~\ref{sec:SUGRA} is accordingly not a change of subject 
but an \emph{elevation of the description}. The Coulomb-branch moduli space 
$\mathcal{C}$, equipped with its rigid special K\"ahler structure, is a particular 
instance of the K\"ahler scalar manifold $\mathcal{M}$ that appears in the general 
$\mathcal{N}=1$ framework. Upon localisation of the supersymmetry parameter, 
$\epsilon_\alpha \to \epsilon_\alpha(x)$, this geometry is no longer an internal 
affair of the field theory: through the curved-superspace measure $E$ and the 
K\"ahler function
\begin{equation}
    G(\varphi,\bar\varphi) 
    = \frac{K(\varphi,\bar\varphi)}{M_{\mathrm{Pl}}^2} 
    + \ln\frac{|W(\varphi)|^2}{M_{\mathrm{Pl}}^6},
\end{equation}
the geometry of $\mathcal{M}$ becomes inextricably coupled to the geometry of 
spacetime. The negative gravitational contribution $-3|W|^2/M_{\mathrm{Pl}}^2$ 
to the scalar potential, absent in the global theory, is precisely the new feature 
that makes the sign of the vacuum energy a dynamical question rather than a 
fixed non-negative number --- and it is this freedom that underlies the uplift 
from a supersymmetric AdS minimum to a metastable de~Sitter vacuum in the 
KKLT construction of Section~\ref{sec:moduli}.

%%%%%%%%%%%%%%%%%%%%%%%%%%%%%%%%%%%%%
\section{Applications to String Theory}\label{sec:strings}

Compactification of the ten-dimensional Type~IIB theory on a Calabi--Yau 
threefold $X_6$ provides a geometric origin for the Kähler potential and 
superpotential that Section~\ref{sec:SUGRA} treated as abstract inputs; the 
moduli of Sections~\ref{sec:SW}--\ref{sec:SUGRA} acquire a direct 
interpretation as coordinates on the compactification space or positions of 
D-branes within it. As demonstrated in Section~\ref{sec:SUGRA}, the localisation of supersymmetry inevitably leads to the inclusion of gravity via the supergravity multiplet. However, supergravity itself is regarded as a low-energy effective limit of a more fundamental theory. In this section, we demonstrate how previously discussed constructions, such as moduli spaces, Kähler geometry, and supersymmetric gauge theories, arise naturally upon compactification of string theories and in the presence of D-branes \cite{Polchinski}. This not only substantiates the physical motivation for the models considered but also places them in the context of the unification of all fundamental interactions.

The transition from the four-dimensional effective field theory to the fundamental string description necessitates the introduction of extra spatial dimensions. This requirement is dictated by several consistency conditions that cannot be satisfied in four dimensions alone:

\begin{enumerate}
    \item \textbf{Anomaly Cancellation:} In the context of quantum gravity, gauge and gravitational anomalies must cancel exactly to preserve unitarity. In superstring theory, this cancellation occurs naturally only in critical dimensions ($D=10$ for superstrings), where the contributions from bosonic and fermionic modes balance precisely.
    
    \item \textbf{Ultraviolet Completion:} While supergravity (Section~\ref{sec:SUGRA}) improves the ultraviolet
behavior of quantum field theory, it remains non-renormalizable at high energies. String theory provides a finite UV completion by replacing point-like interactions with extended objects, thereby resolving the singularities inherent in the local field theory description.
    
    \item \textbf{Unified Origin of Moduli:} The moduli spaces discussed in Sections~\ref{sec:SW}
and~\ref{sec:SUGRA} acquire a direct geometric interpretation in higher dimensions. The scalar fields $\phi^I$ that parameterize the vacuum manifold correspond to the coordinates of compactified extra dimensions and the positions of branes within them.
\end{enumerate}

Upon compactification of the ten-dimensional spacetime on a Calabi--Yau manifold $X_6$, the effective four-dimensional theory inherits the geometric structure of the internal space. The Kähler potential $K(\Phi, \Phi^\dagger)$ and superpotential $W(\Phi)$
introduced in Section~\ref{sec:SUGRA} are determined by the moduli of $X_6$:
\begin{equation}
K \sim -\ln\left(i \int_{X_6} \Omega \wedge \bar{\Omega}\right), \quad W \sim \int_{X_6} G_3 \wedge \Omega,
\end{equation}
where $\Omega$ is the holomorphic $(3,0)$-form of the Calabi--Yau space and $G_3$ represents flux backgrounds. This establishes a direct correspondence between the abstract K\"ahler geometry of the scalar manifold (Section~\ref{sec:SUGRA}) and the actual geometry
of the compactification space.

Furthermore, the gauge theories constructed in Sections~\ref{sec:SYM} --~\ref{sec:SUGRA} emerge naturally from the dynamics of D-branes --- extended objects on which open strings can end. The worldvolume theory on a stack of $N$ coincident D3-branes reproduces the $\mathcal{N}=4$ supersymmetric Yang--Mills action, while configurations at singularities or intersections yield $\mathcal{N}=2$ and $\mathcal{N}=1$ theories with matter content. This provides a geometric origin for the gauge symmetries, matter representations, and superpotential couplings that were introduced axiomatically in the field-theoretic framework.

\subsection{D-Branes and Supersymmetric Gauge Theories}
\label{sec:6.2}

Supersymmetric gauge theories arise naturally on the world volume of D-branes~\cite{Polchinski}. This provides a geometric origin for the gauge symmetries and matter content discussed in the previous sections. Table~\ref{tab:dbrane_gauge} summarizes the correspondence between D-brane configurations and the resulting supersymmetric gauge theories on their worldvolume. The key observations are:

\begin{itemize}
    \item \textbf{Maximal supersymmetry:} A stack of $N$ coincident D3-branes in flat ten-dimensional spacetime preserves all 16 supercharges, yielding $\mathcal{N}=4$ supersymmetric Yang--Mills theory with gauge group $U(N)$. This is the maximally supersymmetric gauge theory in four dimensions, which was introduced in our general discussion of extended supersymmetry.
    
    \item \textbf{Reduced supersymmetry via geometry:} Placing D3-branes at singularities of the compactification space (such as orbifold or conifold singularities) breaks part of the supersymmetry. Depending on the type of singularity, one can engineer theories with $\mathcal{N}=2$ or $\mathcal{N}=1$ supersymmetry, corresponding to 8 or 4 preserved supercharges respectively.
    
    \item \textbf{Matter from additional branes:} The D3/D7 system introduces fundamental matter (flavors) into the gauge theory. The D7-branes extend in additional directions compared to the D3-branes, and open strings stretching between D3 and D7 branes give rise to hypermultiplets in the fundamental representation. This provides a geometric realization of the matter content discussed in Sections~2--3.
    
    \item \textbf{Chiral theories from intersections:} Intersecting D-brane configurations at generic angles preserve only $\mathcal{N}=1$ supersymmetry (4 supercharges) and naturally produce chiral matter --- a crucial ingredient for constructing realistic particle physics models. The chirality arises from the specific intersection geometry and the GSO projection of open string states.
\end{itemize}

This classification demonstrates how the abstract supersymmetric field theories discussed throughout this review emerge from concrete geometric configurations in string theory. The amount of preserved supersymmetry is directly controlled by the brane configuration and the geometry of the extra dimensions, providing a powerful framework for engineering quantum field theories with desired properties as presented in Table ~\ref{tab:dbrane_gauge}.

\begin{table}[ht]
\centering
\caption{Gauge theories from D-brane configurations.}
\begin{tabular}{|l|l|l|}
\hline
\textbf{D-Brane} & \textbf{Worldvolume Theory} & \textbf{Supersymmetry} \\
\hline
D3-brane & $ \mathcal{N}=4 $ SYM & 16 supercharges \\
D3 at singularity & $ \mathcal{N}=2 $ or $ \mathcal{N}=1 $ & 8 or 4 supercharges \\
D3/D7 system & $ \mathcal{N}=2 $ with flavors & 8 supercharges \\
Intersecting D-branes & $ \mathcal{N}=1 $ chiral matter & 4 supercharges \\
\hline
\end{tabular}
\label{tab:dbrane_gauge}
\end{table}

In particular, $ \mathcal{N}=4 $ supersymmetric Yang--Mills theory appears as the low energy limit of open strings attached to D3-branes. The action takes the form:

\begin{equation}
S_{\mathcal{N}=4} = \frac{1}{g_{\text{YM}}^2} \int d^4x \, \text{Tr} \left[ -\frac{1}{4} F_{\mu\nu} F^{\mu\nu} + \frac{1}{2} D_\mu \phi^I D^\mu \phi^I + \frac{1}{4} [\phi^I, \phi^J]^2 + \text{fermions} \right],
\label{eq:n4_sym_action}
\end{equation}
where $ \phi^I $ ($ I = 1,\ldots,6 $) are the six adjoint scalars corresponding to transverse fluctuations of the D3-brane. The structure of action~\eqref{eq:n4_sym_action} directly reflects the supersymmetric components developed in the preceding sections. 
The gauge kinetic term $\frac{1}{4} F_{\mu\nu}F^{\mu\nu}$ corresponds to the vector multiplet dynamics introduced in Section 2.2 (Eq. 12). 
The six scalar fields $\phi^I$ generalize the chiral multiplet scalars discussed in the Wess--Zumino model (Section 3) and the moduli space coordinates of the Coulomb branch analyzed in Section 4.2. 
The scalar potential term $\frac{1}{4}[\phi^I, \phi^J]^2$ in ~\eqref{eq:n4_sym_action} in $\mathcal{N}=4$ SYM can be derived by decomposing the theory into 
$\mathcal{N}=1$ superfields: one vector multiplet $V$ and three chiral multiplets $\Phi_i$ 
($i=1,2,3$) in the adjoint representation. The superpotential takes the form
\begin{equation}
W = \frac{g}{3} \epsilon_{ijk} \text{Tr}(\Phi^i \Phi^j \Phi^k).
\end{equation}
The F-term contribution is
\begin{equation}
V_F = \sum_i \text{Tr}\left|\frac{\partial W}{\partial \phi_i}\right|^2 
= \frac{g^2}{4} \sum_{i,j} \text{Tr}\left|[\phi^i, \phi^j]\right|^2,
\end{equation}
while the D-term contribution for adjoint fields is
\begin{equation}
V_D = \frac{g^2}{2} \sum_a \left(\sum_i \phi_i^\dagger T^a \phi_i\right)^2
= \frac{g^2}{4} \sum_{i,j} \text{Tr}\left([\phi_i^\dagger, \phi_j^\dagger][\phi_i, \phi_j]\right).
\end{equation}
Combining the six real scalars $\phi^I$ ($I=1,\ldots,6$) from the three complex fields $\phi^i$, 
the $\mathcal{N}=4$ supersymmetry ensures that $V_F$ and $V_D$ merge into the compact commutator form:
\begin{equation}
V = V_F + V_D = \frac{1}{4} \sum_{I<J} \text{Tr}\left[\phi^I, \phi^J\right]^2.
\label{eq:scalar}
\end{equation}
The vanishing condition $[\phi^I, \phi^J]=0$ defines the moduli space where all scalars can be 
simultaneously diagonalized, corresponding to separated D-branes in the transverse geometry.

Physically, these scalars represent the transverse fluctuations of the D3-brane in the ten-dimensional spacetime. 
Thus, the geometric moduli space of the gauge theory (Sections 4--5) acquires a direct spatial interpretation as the position of the brane in the extra dimensions. 
This confirms that the effective field theory formalism constructed in Sections 2--5 emerges naturally from the fundamental string theoretic objects.

\subsection{Gauge/Gravity Duality}

The D-brane construction provides the foundation for gauge/gravity duality (AdS/CFT correspondence)~\cite{Maldacena1998}. The key insight is that D-branes admit two complementary descriptions:

\begin{enumerate}
    \item \textbf{Open string description:} Gauge theory on the brane worldvolume (Eq.~\eqref{eq:n4_sym_action})
    \item \textbf{Closed string description:} Supergravity in the near-horizon geometry
\end{enumerate}

For a stack of $N$ D3-branes, these descriptions are related by the AdS/CFT correspondence:
\begin{equation}
\boxed{\mathcal{N}=4 \text{ SYM with gauge group SU}(N) \quad \Longleftrightarrow \quad \text{Type IIB SUGRA on AdS}_5 \times S^5}
\label{eq:70}
\end{equation}

The parameters are related as:
\begin{equation}
g_{\text{YM}}^2 = 4\pi g_s, \qquad \lambda = g_{\text{YM}}^2 N = \frac{R^4}{\alpha'^2},
\label{eq:71}
\end{equation}
where:
\begin{itemize}
    \item $\mathcal{N}=4$ denotes the number of supersymmetries (maximal supersymmetry in four dimensions, corresponding to 16 supercharges);
    \item $\text{SU}(N)$ is the special unitary gauge group of rank $N-1$, where $N$ represents the number of coincident D3-branes;
    \item Type IIB SUGRA refers to Type IIB supergravity, the low-energy effective field theory of Type IIB superstring theory;
    \item $\text{AdS}_5$ is five-dimensional anti-de Sitter space with radius $R$, representing the near-horizon geometry of the D3-brane stack;
    \item $S^5$ is the five-dimensional sphere of radius $R$, arising from the transverse space to the D3-branes;
    \item $g_{\text{YM}}$ is the Yang--Mills coupling constant of the gauge theory;
    \item $g_s$ is the string coupling constant, controlling the strength of string interactions;
    \item $\lambda = g_{\text{YM}}^2 N$ is the 't Hooft coupling, which remains fixed in the large-$N$ limit;
    \item $R$ is the common radius of both $\text{AdS}_5$ and $S^5$;
    \item $\alpha' = \ell_s^2$ is the Regge slope parameter, where $\ell_s$ is the string length scale.
\end{itemize}

The relation $\lambda = R^4/\alpha'^2$ implies that the strong coupling limit $\lambda \gg 1$ of the gauge theory corresponds to the classical supergravity regime $R \gg \ell_s$, where curvature corrections are suppressed. Conversely, weak coupling $\lambda \ll 1$ corresponds to the stringy regime where $\alpha'$ corrections become important.

\subsection{Connection to Seiberg--Witten Theory}
\label{sec:6.3}

The geometric engineering of gauge theories via D-branes provides a string-theoretic realization of Seiberg--Witten theory~\cite{Klemm:1996bj}. The correspondence between the brane picture and the field-theoretic description can be summarized as follows:

\begin{itemize}
    \item \textbf{Coulomb branch:} Corresponds to positions of $N$ D3-branes in the transverse $\mathbb{R}^6$ space. For $SU(2)$ gauge theory, two D3-branes separated by distance $2a$ in the transverse directions break the gauge symmetry $U(2) \to U(1) \times U(1)$. The gauge-invariant coordinate is
    \begin{equation}
        u = \text{Tr}\,\phi^2 = 2a^2,
    \end{equation}
    where $\phi$ is the scalar field in the vector multiplet, and $a$ represents the brane separation.
    
    \item \textbf{Monodromy:} Arises from brane motion around singularities. When D3-branes coincide ($u \to \Lambda^2$), additional states become massless. Encircling this point in the moduli space induces an $SL(2,\mathbb{Z})$ transformation on the period vector $(a_D, a)$:
    \begin{equation}
        \begin{pmatrix} a_D \\ a \end{pmatrix} \longrightarrow 
        \begin{pmatrix} 1 & 0 \\ -1 & 1 \end{pmatrix}
        \begin{pmatrix} a_D \\ a \end{pmatrix}.
    \end{equation}
    This monodromy reflects the exchange of branch points on the Seiberg--Witten curve.
    
    \item \textbf{BPS states:} Realized as fundamental strings (F1) and D1-branes stretching between D3-branes. A string stretched between two D3-branes separated by distance $2a$ has mass
    \begin{equation}
        M = T_{\text{F1}} \cdot 2a = \frac{2a}{2\pi\alpha'} = |n_e a|,
    \end{equation}
    where $n_e$ is the electric charge. Magnetic monopoles correspond to D1-branes ending on D3-branes, with mass $M = |n_m a_D|$. General $(p,q)$ strings give dyonic states with mass
    \begin{equation}
        M_{(p,q)} = |p a + q a_D|,
    \end{equation}
    reproducing the exact BPS mass formula from Seiberg--Witten theory.
    
    \item \textbf{Elliptic curve:} Emerges from the geometry of the brane configuration. For $SU(2)$ theory, the Seiberg--Witten curve
    \begin{equation}
        y^2 = (x - e_1(u))(x - e_2(u))(x - e_3(u))
    \end{equation}
    describes an elliptic fibration over the $u$-plane. The branch points $e_i(u)$ collide at singular loci $u = \pm\Lambda^2$, corresponding to coincident branes.
\end{itemize}

\begin{figure}[t]
\centering
\begin{tikzpicture}[scale=0.9]

% LEFT: Separated D3-branes
\draw[thick, blue, fill=blue!20] (0,4) rectangle (4,4.3);
\draw[thick, blue, fill=blue!20] (0,2) rectangle (4,2.3);
\draw[ultra thick, red] (2,2.3) -- (2,4);
\draw[<->] (4.3,4) -- (4.3,2.3) node[midway,right] {$2a$};
\node[above] at (2,4.3) {D3-brane 1};
\node[below] at (2,2) {D3-brane 2};
\node[right, red] at (2.1,3.2) {F1 string};

% RIGHT: Coincident D3-branes
\draw[ultra thick, red, fill=red!20] (6,3) rectangle (10,3.5);
\filldraw[yellow] (6.5,3.25) circle (3pt);
\filldraw[yellow] (8,3.25) circle (3pt);
\filldraw[yellow] (9.5,3.25) circle (3pt);
\draw[thick, magenta] (8,3.25) ellipse (1 and 0.5);
\node[above] at (8,3.5) {Coincident D3};
\node[below, red] at (8,2.5) {$u = \Lambda^2$};

% BOTTOM: u-plane
\draw[->] (-1,-1) -- (5,-1) node[right] {Re$(u)$};
\draw[->] (2,-2.5) -- (2,0.5) node[above] {Im$(u)$};
\filldraw[red] (0.5,-1) circle (3pt) node[below] {$-\Lambda^2$};
\filldraw[red] (3.5,-1) circle (3pt) node[below] {$\Lambda^2$};
\draw[blue, thick] (3.5,-1) ellipse (0.6 and 0.4);
\node[blue, right] at (4.2,-1) {$\gamma$};

% Titles
\node[font=\bfseries] at (2,5) {Separated ($u \neq \Lambda^2$)};
\node[font=\bfseries] at (8,5) {Coincident ($u = \Lambda^2$)};
\node[font=\bfseries] at (2,-2) {$u$-plane};

\end{tikzpicture}
\caption{D-brane realization of Seiberg--Witten theory. Left: separated D3-branes with F1 string. Right: coincident branes at singularity with massless BPS states. Bottom: $u$-plane with singularities and monodromy loop $\gamma$.}
\label{fig:dbrane_sw}
\end{figure}
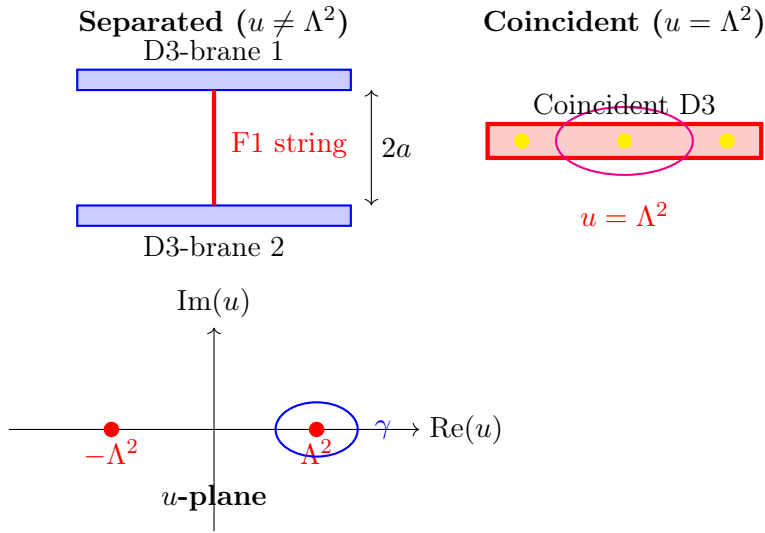

The Seiberg--Witten differential $\lambda_{\text{SW}}$ can be derived from the period integrals of the holomorphic three-form $\Omega_3$ on the Calabi--Yau threefold in Type IIB compactifications:
\begin{equation}
    a = \oint_{\alpha} \lambda_{\text{SW}}, \quad a_D = \oint_{\beta} \lambda_{\text{SW}}, \quad \text{where} \quad \lambda_{\text{SW}} = \int_{\Sigma_2} \Omega_3.
\end{equation}
Here $\Sigma_2$ is a two-cycle in the internal geometry that connects the D-brane positions.

This brane picture provides an intuitive geometric understanding of the non-perturbative dynamics: quantum corrections deform the classical moduli space, introducing singularities where the effective description breaks down and new light degrees of freedom must be included. The duality group $SL(2,\mathbb{Z})$ emerges naturally as the mapping class group of the torus fiber in the elliptic fibration.

\subsection{From \texorpdfstring{$\mathcal{N}=4$}{N=4} to \texorpdfstring{$\mathcal{N}=1$}{N=1}: Supersymmetry Breaking}

Phenomenologically viable models require $\mathcal{N}=1$ supersymmetry, since the
maximal $\mathcal{N}=4$ theory is far too constrained to accommodate chiral matter,
realistic Yukawa couplings, or spontaneous symmetry breaking patterns compatible with
the Standard Model.  The mechanisms that connect the maximally supersymmetric theory
to $\mathcal{N}=1$ supergravity fall into three conceptually distinct classes, each of
which modifies the superpotential $W$ and the K\"{a}hler potential $K$ in a
characteristic way.  We discuss them in increasing order of their structural
complexity.

\paragraph{Orbifold Projections.}

The most direct geometric mechanism is to quotient the internal space by a discrete
symmetry group $\Gamma$, a procedure known as an orbifold projection.  When Type~IIB
string theory is compactified on the orbifold $\mathbb{C}^3/\mathbb{Z}_3$, only one
of the original four supercharges survives the projection, reducing $\mathcal{N}=4$
to $\mathcal{N}=1$.  At the field-theory level, the projection is implemented by
retaining only those fields that are invariant under the combined action of $\Gamma$
on the gauge and R-symmetry indices:
\begin{equation}\label{eq:orbifold_proj}
  \Phi \;\longmapsto\; P\Phi
  \;=\; \frac{1}{|\Gamma|}\sum_{g\in\Gamma} \gamma_g\,\Phi\,\gamma_g^{-1},
\end{equation}
where $\gamma_g$ represents the action of the group element $g$ on the gauge indices.
The projected theory has a gauge group of the form $\prod_i U(N_i)$, whose quiver
structure is entirely determined by the embedding of $\Gamma$ into the
$SU(4)_R$ R-symmetry group of the parent $\mathcal{N}=4$ theory.  The matter content
--- the set of bifundamental chiral multiplets connecting adjacent nodes of the quiver
--- follows from the decomposition of the adjoint representation of $\mathcal{N}=4$
under the orbifold action.

\paragraph{Mass Deformations.}

A complementary, purely field-theoretic route to $\mathcal{N}=1$ is to introduce
explicit mass terms for some of the adjoint chiral multiplets of the parent theory.
The $\mathcal{N}=4$ superpotential can be written in $\mathcal{N}=1$ language as a
cubic coupling of three adjoint chiral superfields $\Phi^i$:
\begin{equation}\label{eq:W_N4}
  W_{\mathcal{N}=4} \;=\; \frac{g}{3}\,\epsilon_{ijk}\,
  \operatorname{Tr}\!\left(\Phi^i\Phi^j\Phi^k\right).
\end{equation}
Adding mass terms for the adjoint multiplets,
\begin{equation}\label{eq:delta_W}
  \Delta W \;=\; \frac{1}{2}\sum_{i=1}^{3} m_i\,
  \operatorname{Tr}\!\left(\Phi^i\Phi^i\right),
\end{equation}
deforms the theory in a way that depends sensitively on the mass ratios.  If only one
mass is taken to infinity while the other two vanish, the light degrees of freedom
constitute an $\mathcal{N}=2$ theory; if two masses are sent to infinity, one is left
with pure $\mathcal{N}=1$ Yang--Mills; and if all three masses are equal and
nonzero, the theory flows to the $\mathcal{N}=1^*$ theory studied by Donagi and
Witten.  In every case the deformation lifts the flat directions of the Coulomb branch
by adding terms proportional to $|m_i\phi^i|^2$ to the scalar potential $\Vtot$,
driving the vacuum away from the region where the $\mathcal{N}=4$ moduli space
analysis of Section~3 applies.

\paragraph{Flux Compactifications and the Role of the Dilaton.}

The most natural and physically motivated mechanism, however, arises from the
geometry of the compactification itself.  When the ten-dimensional Type~IIB theory is
compactified on a Calabi--Yau threefold $X_6$, one may thread quantised fluxes of the
Ramond--Ramond field strength $F_3$ and the Neveu--Schwarz three-form $H_3$ through
the internal cycles.  These fluxes generate a superpotential for the moduli of $X_6$
via the Gukov--Vafa--Witten formula:
\begin{equation}\label{eq:W_flux}
  W_{\mathrm{flux}} \;=\; \int_{X_6} G_3 \wedge \Omega,
  \qquad G_3 \;=\; F_3 - \tau H_3,
\end{equation}
where $\Omega$ is the holomorphic $(3,0)$-form of $X_6$ and $\tau$ is the
axion--dilaton modulus.  The latter combines the Ramond--Ramond axion $C_0$ with the
dilaton $\phi$ into the complex field
\begin{equation}\label{eq:tau_dilaton}
  \tau \;=\; C_0 + i\,e^{-\phi} \;=\; C_0 + \frac{i}{g_s},
\end{equation}
which transforms as a modular parameter under the $SL(2,\mathbb{Z})$ S-duality group
of Type~IIB string theory.  In the four-dimensional effective supergravity, $\tau$ is
promoted to a chiral superfield $S$ whose real part governs the gauge coupling of the
hidden-sector gauge group, $\operatorname{Re}(S) = 1/g_{\mathrm{YM}}^2$.  This
relation is the reason the dilaton must be stabilised: since it sets the scale of all
non-perturbative effects, a runaway in the dilaton direction would render the entire
vacuum structure meaningless.

Supersymmetry is preserved by the fluxes if and only if $G_3$ is imaginary self-dual,
$*_6 G_3 = iG_3$.  Deviating from this condition generates a non-vanishing $F$-term
for the moduli and spontaneously breaks supersymmetry, giving the complex-structure
moduli and the dilaton masses of order the flux scale.  The K\"{a}hler potential
acquires a contribution from the volume modulus $\mathcal{V}$ of the internal space:
\begin{equation}\label{eq:K_flux}
  K \;=\; -\ln\!\left(-i\!\int_{X_6}\!\Omega\wedge\bar{\Omega}\right)
          - 2\ln\mathcal{V} + K_{\mathrm{matter}}.
\end{equation}
A crucial observation --- sometimes called the no-scale structure of flux
compactifications --- is that while $W_{\mathrm{flux}}$ fixes the complex-structure
moduli and $\tau$ at a supersymmetric minimum with $D_i W = 0$, it leaves the
K\"{a}hler modulus $\mathcal{V}$ (or equivalently the volume modulus
$\mathcal{T} = \operatorname{Re}\mathcal{T} + i\,b$) entirely undetermined at this
stage.  Stabilising the remaining modulus requires non-perturbative effects.

\paragraph{Gaugino Condensation and Moduli Stabilisation.}

Non-perturbative effects in a hidden-sector gauge group with $N$ colours generate
a gaugino condensate $\langle\lambda\lambda\rangle \neq 0$ below a dynamical scale
$\Lambda_{\mathrm{np}}$.  In the low-energy supergravity description this condensate
manifests itself as a non-perturbative contribution to the superpotential:
\begin{equation}\label{eq:W_np}
  W_{\mathrm{np}} \;=\; A\,e^{-aS}, \qquad a \;=\; \frac{8\pi^2}{N},
\end{equation}
where $A$ is a one-loop determinant factor and $N$ is the rank of the condensing
gauge group.  Physically, this term arises because the gauge coupling
$g_{\mathrm{YM}}^2 \propto 1/\operatorname{Re}(S)$ runs with the renormalisation
group, and at the scale where it becomes of order unity instantons (or equivalently
the gaugino condensate) produce a non-perturbative correction to the holomorphic
functions $K$ and $W$.  Combining the flux superpotential with the non-perturbative
term gives the KKLT superpotential \cite{KKLT2003},
\begin{equation}\label{eq:W_KKLT}
  W \;=\; W_{\mathrm{flux}} + W_{\mathrm{np}}
    \;=\; W_0 + A\,e^{-a\mathcal{T}},
\end{equation}
where $W_0 = \langle W_{\mathrm{flux}}\rangle$ is the constant value of the flux
superpotential at the minimum for the complex-structure moduli, and $\mathcal{T}$ is
the K\"{a}hler modulus.  The competition between the constant $W_0$ (which tends to
push the potential to negative values) and the exponential $Ae^{-a\mathcal{T}}$
(which is positive and falls off at large volume) creates a supersymmetric AdS minimum
at a finite value $\mathcal{T}_0$ determined by $D_{\mathcal{T}}W = 0$.

\paragraph{Vacuum Structure and the Uplift to de~Sitter Space.}

The effective scalar potential that emerges from this construction matches the
general $\mathcal{N}=1$ SUGRA form discussed in Section~\ref{sec:SUGRA}:
\begin{equation}\label{eq:V_SUGRA_full}
  V = M_{\mathrm{Pl}}^{4}\,e^{G}\!\left(g^{i\bar{\jmath}}G_{i}G_{\bar{\jmath}}
      - 3\right), \qquad
  G = \frac{K}{M_{\mathrm{Pl}}^{2}}
    + \ln\frac{|W|^{2}}{M_{\mathrm{Pl}}^{6}}.
\end{equation}
Depending on the interplay between the flux terms and the non-perturbative
corrections, three qualitatively distinct vacuum types are realised.

%------------------------------------------------------------------
% STAGE 1
%------------------------------------------------------------------
\medskip
\noindent\textit{Stage~1: Supersymmetric Minkowski Vacuum.}
In the first stage the system resides in a supersymmetric Minkowski vacuum,
characterised by the simultaneous vanishing of both the K\"{a}hler-covariant
derivative of the superpotential and the superpotential itself,
\begin{equation}\label{eq:stage1}
  D_{i}W = 0, \quad W = 0 \quad\Longrightarrow\quad \Lambda = 0.
\end{equation}
Under these conditions all $F$-terms vanish identically,
\begin{equation}
  F^{i} = e^{K/2}\,K^{i\bar{\jmath}}\,D_{\bar{\jmath}}\bar{W} = 0,
\end{equation}
and the scalar potential attains its minimum value $V_{\min}=0$, yielding a
perfectly flat Minkowski spacetime. Supersymmetry remains unbroken, so that all
superpartners are mass-degenerate, $m_{\mathrm{boson}}=m_{\mathrm{fermion}}$,
and no spontaneous SUSY breaking occurs. The vanishing of the cosmological
constant follows as a direct consequence of~\eqref{eq:stage1}, rather than as
an independent assumption.

%------------------------------------------------------------------
% STAGE 2
%------------------------------------------------------------------
\medskip
\noindent\textit{Stage~2: Flux Stabilisation and the Supersymmetric AdS Vacuum.}
The second stage introduces flux stabilisation, leading to a supersymmetric
Anti-de~Sitter vacuum. The condition $D_{i}W=0$ is preserved, but a small
nonzero flux superpotential $W_{0}\neq 0$ is generated by the Gukov--Vafa--Witten
mechanism~\eqref{eq:W_flux}, rendering the cosmological constant negative,
$\Lambda < 0$. This is precisely the KKLT minimum before uplifting: all moduli
are stabilised by the competition between the constant $W_{0}$ and the
non-perturbative term $Ae^{-a\mathcal{T}}$ in~\eqref{eq:W_KKLT}, but the vacuum
energy is negative. While the complex-structure moduli and the dilaton are fixed
by the flux-induced superpotential, the K\"{a}hler modulus $\mathcal{V}$ remains
unstabilised at the purely perturbative level, a well-known shortcoming that
motivates the gaugino condensation mechanism discussed above and the subsequent
uplift procedure.

%------------------------------------------------------------------
% STAGE 3
%------------------------------------------------------------------
\medskip
\noindent\textit{Stage~3: Uplift to a Metastable de~Sitter Vacuum.}
The third stage achieves an uplift to a metastable de~Sitter vacuum through the
inclusion of $\overline{\mathrm{D3}}$-branes, whose positive-energy contribution
to the scalar potential takes the form
\begin{equation}\label{eq:V_D3bar}
  V_{\overline{\mathrm{D3}}} = \frac{D}{(\operatorname{Re}\mathcal{T})^{3}},
  \qquad D > 0.
\end{equation}
The total scalar potential therefore reads
\begin{equation}\label{eq:V_total}
  V_{\mathrm{total}} = V_{F\text{-term}} + V_{\overline{\mathrm{D3}}},
\end{equation}
where the $F$-term contribution is
\begin{equation}\label{eq:VF_KKLT}
  V_{F} = e^{K}\!\left(K^{T\bar{T}}\,D_{T}W\,D_{\bar{T}}\bar{W}
          - 3\lvert W\rvert^{2}\right),
\end{equation}
and $D_{T}W = \partial_{T}W + (\partial_{T}K)\,W$ denotes the
K\"{a}hler-covariant derivative. The $\overline{\mathrm{D3}}$-brane contribution
explicitly breaks supersymmetry and generates a small positive cosmological
constant $\Lambda > 0$, while all moduli acquire stabilised masses. The resulting
vacuum is metastable with lifetime $\tau\gg t_{\mathrm{Hubble}}$, providing a
concrete realisation of the KKLT mechanism for explaining the observed dark
energy, for constructing inflationary models based on K\"{a}hler moduli, and for
situating the string landscape within an anthropic framework.

The soft supersymmetry-breaking terms communicated to the visible sector through
gravity mediation are characterised by the scale
\begin{equation}\label{eq:msoft}
  m_{\mathrm{soft}} \sim \frac{F}{M_{\mathrm{Pl}}},
  \qquad
  F_{i} = e^{K/2}\,K^{i\bar{\jmath}}\,D_{\bar{\jmath}}\bar{W},
\end{equation}
which connects the geometric breaking mechanism in the hidden sector to the
phenomenological mass spectrum of the observable sector.

Taken together, this three-stage construction --- supersymmetric Minkowski space,
flux-induced AdS stabilisation, and $\overline{\mathrm{D3}}$-brane uplift ---
illustrates how D-brane gauge theories, Seiberg--Witten dynamics, and
$\mathcal{N}=1$ supergravity combine into a unified framework for analysing
non-perturbative quantum field theory and its phenomenological applications to
particle physics and cosmology.

\begin{figure}[ht]
\centering
\begin{tikzpicture}[node distance=1.5cm, auto]
    \node (string) [rectangle, draw, fill=blue!10] {String Theory};
    \node (dbrane) [rectangle, draw, below of=string] {D-Branes};
    \node (gauge) [rectangle, draw, left of=dbrane, xshift=-2cm] {Gauge Theory};
    \node (sugra) [rectangle, draw, right of=dbrane, xshift=2cm] {Supergravity};
    \node (sw) [rectangle, draw, below of=dbrane] {Seiberg--Witten};
    \node (phenom) [rectangle, draw, fill=green!10, below of=sw] {Phenomenology};
    
    \draw[->] (string) -- (dbrane);
    \draw[->] (dbrane) -- (gauge);
    \draw[->] (dbrane) -- (sugra);
    \draw[->] (gauge) -- (sw);
    \draw[->] (sugra) -- (sw);
    \draw[->] (sw) -- (phenom);
    \draw[->, dashed] (sugra) -- (phenom);
\end{tikzpicture}
\caption{Conceptual connections between string theory, gauge theories, supergravity, and phenomenology.}
\label{fig:string_connections}
\end{figure}
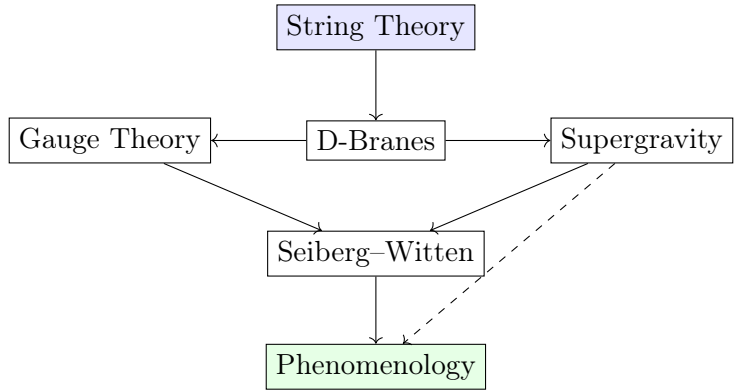
Figure~\ref{fig:string_connections} summarizes the web of connections established throughout this work. The D-brane construction serves as the bridge between fundamental string theory and effective field theory descriptions, while Seiberg--Witten theory and supergravity provide complementary tools for analyzing non-perturbative dynamics and vacuum structure.

\subsection{From Abstract Supergravity to Geometric Realisation}
\label{subsec:sec4_sec5_connection}

Section~\ref{sec:SUGRA} concludes with a central structural observation: supergravity is not merely
the globally supersymmetric theory with gravitational interactions appended, but a
genuinely new framework in which the geometry of spacetime and the geometry of the
scalar field manifold $\mathcal{M}$ become inextricably coupled through three
holomorphic functions --- the K\"{a}hler potential $K(\Phi_i, \Phi^\dagger_i)$,
the superpotential $W(\Phi_i)$, and the gauge kinetic function $f_{ab}(\Phi_i)$.
Once these functions are specified, every sector of the Lagrangian is uniquely
determined; no further freedom remains. This raises an immediate and fundamental
question: \emph{what determines $K$, $W$, and $f_{ab}$ themselves?}

Within the purely field-theoretic framework of Section~\ref{sec:SUGRA}, these functions are
postulated as abstract inputs. Section~\ref{sec:strings} answers the question geometrically:
in the context of superstring theory compactified on a Calabi--Yau threefold
$X_6$, the holomorphic functions $K$, $W$, and $f_{ab}$ are \emph{computed}
from the geometry of the internal space and the dynamics of D-branes. The
transition from Section~\ref{sec:SUGRA} to Section~\ref{sec:strings} is therefore not a change of subject
but an elevation of the level of description --- from an abstract effective
field theory to its geometric and string-theoretic origin.

\subsubsection*{Geometric Origin of the K\"{a}hler Potential and Superpotential}

In Section~\ref{sec:SUGRA}, the K\"{a}hler potential is introduced axiomatically as a real
function whose second derivatives define the scalar metric,
$g_{i\bar{\jmath}} = \partial_i \partial_{\bar{\jmath}} K$.
Upon compactification of Type~IIB string theory on a Calabi--Yau threefold
$X_6$, this function is determined explicitly by the geometry of the internal
space~\cite{Polchinski}:
\begin{equation}
    K \;\sim\; -\ln\!\left( i \int_{X_6} \Omega \wedge \bar{\Omega} \right)
    - 2\ln \mathcal{V} + K_{\mathrm{matter}},
\label{eq:Kahler_CY}
\end{equation}
where $\Omega$ is the holomorphic $(3,0)$-form of the Calabi--Yau manifold and
$\mathcal{V}$ is the volume of the internal space measured in string units.
Likewise, the superpotential, which in Section~\ref{sec:SUGRA} appears as an abstract
holomorphic function of the chiral superfields, acquires a concrete realisation
through the Gukov--Vafa--Witten formula~\cite{Gukov:1999ya}:
\begin{equation}
    W \;\sim\; \int_{X_6} G_3 \wedge \Omega,
    \qquad
    G_3 = F_3 - \tau H_3,
\label{eq:GVW}
\end{equation}
where $F_3$ and $H_3$ are the quantised Ramond--Ramond and Neveu--Schwarz
three-form fluxes threading the internal cycles, and
$\tau = C_0 + i\,e^{-\phi} = C_0 + i/g_s$
is the axion--dilaton modulus. The gauge kinetic function correspondingly takes
the form $f_{ab} \sim \tau\,\delta_{ab}$, so that the real part of $\tau$
determines the gauge coupling and its imaginary part the topological
$\theta$-angle of the effective theory.

The three holomorphic functions that Section~4 treats as free inputs are thus
fixed by the geometry of $X_6$ and the choice of quantised flux background.
Table~\ref{tab:sec4_sec5} summarises the precise correspondence between the
abstract supergravity data of Section~\ref{sec:SUGRA} and their string-theoretic origins
in Section~\ref{sec:strings}.

\begin{table}[h]
\centering
\caption{Correspondence between the abstract $\mathcal{N}=1$ supergravity data
of Section~4 and their geometric origins in the string-theoretic framework of
Section~5. The holomorphic functions $K$, $W$, and $f_{ab}$, which are free
inputs in the effective field theory, are determined by the compactification
geometry and the flux background.}
\renewcommand{\arraystretch}{1}
\begin{tabular}{lll}
\hline
\textbf{Object} 
    & \textbf{Section~\ref{sec:SUGRA}: abstract supergravity} 
    & \textbf{Section~\ref{sec:strings}: geometric origin} \\
\hline
K\"{a}hler potential $K$
    & postul. real funct. of $\Phi_i$, $\Phi^\dagger_i$
    & Eq.~\eqref{eq:Kahler_CY}, from geometry of $X_6$ \\
Superpotential $W$
    & postul. holomorphic funct. of $\Phi_i$
    & Eq.~\eqref{eq:GVW}, GVW formula \\
Gauge kinetic funct. $f_{ab}$
    & postul. holomorphic funct. of $\Phi_i$
    & axion--dilaton $\tau = C_0 + i/g_s$ \\
Scalar manifold $\mathcal{M}$
    & abstract K\"{a}hler manifold
    & moduli space of $X_6$ \\
Gauge group $G$
    & postulated
    & worldvolume theory of D-branes \\
Matter representations
    & postulated
    & open strings at brane intersect. \\
Gravitino mass $m_{3/2}$
    & $e^{K/2M_{\mathrm{Pl}}^2}|W|/M_{\mathrm{Pl}}^2$
    & flux scale and gaugino cond. \\
\hline
\end{tabular}
\label{tab:sec4_sec5}
\end{table}

\subsubsection*{D-Branes as the Geometric Origin of Gauge Theories}

A second and equally fundamental connection concerns the origin of the gauge
symmetries and matter content. In the field-theoretic framework of Sections~\ref{sec:SYM}--~\ref{sec:SUGRA},
the gauge group $G$, the matter representations, and the superpotential couplings
are introduced by hand. Section~\ref{sec:strings} provides their geometric origin: gauge theories
arise naturally on the worldvolume of D-branes, with the amount of preserved
supersymmetry controlled entirely by the brane configuration and the geometry
of the extra dimensions.

A stack of $N$ coincident D3-branes in flat ten-dimensional spacetime preserves
all sixteen supercharges and carries $\mathcal{N}=4$ super Yang--Mills with
gauge group $U(N)$ on its worldvolume. Reducing to phenomenologically viable
$\mathcal{N}=1$ models requires breaking part of this supersymmetry, which
can be achieved by three structurally distinct mechanisms: orbifold projections
of the internal geometry, explicit mass deformations of the adjoint chiral
multiplets, and the threading of quantised $G_3$ fluxes through the cycles of
the Calabi--Yau. In each case the K\"{a}hler potential and superpotential of
the resulting four-dimensional effective theory are determined by the compactification
geometry, giving a concrete geometric realisation of the abstract field-theoretic
structures constructed in Section~\ref{sec:SUGRA}.

\subsubsection*{The Scalar Potential and Moduli Stabilisation}

The most direct and consequential link between Sections~\ref{sec:SUGRA} and ~\ref{sec:strings}  concerns the
scalar potential. In Section~\ref{sec:SUGRA} the general $\mathcal{N}=1$ supergravity scalar
potential takes the form
\begin{equation}
    V = e^{K/M_{\mathrm{Pl}}^2}
    \!\left(
        K^{i\bar{\jmath}}\, D_i W\, \overline{D_j W}
        - \frac{3}{M_{\mathrm{Pl}}^2} |W|^2
    \right)
    + \frac{1}{2}(\mathrm{Re}\,f)^{-1\,ab} D_a D_b,
\label{eq:sugra_potential}
\end{equation}
where $D_i W = \partial_i W + (\partial_i K / M_{\mathrm{Pl}}^2)\, W$ is the
K\"{a}hler-covariant derivative. This expression is abstract: it holds for
any choice of $K$ and $W$.

Section~\ref{sec:strings} populates \eqref{eq:sugra_potential} with specific content. The
flux-generated superpotential \eqref{eq:GVW} fixes the complex-structure
moduli and the dilaton at a supersymmetric locus $D_i W = 0$, leaving the
K\"{a}hler modulus $T = \sigma + i\theta$ as a flat direction --- the
no-scale structure identified in Section~\ref{sec:moduli}. The non-perturbative contribution

\begin{equation}
    W_{\mathrm{np}} = A\, e^{-aT},
\label{eq:Wnp}
\end{equation}
arising from gaugino condensation or Euclidean D3-brane instantons in the
hidden-sector gauge group, breaks this degeneracy and generates through
\eqref{eq:sugra_potential} a non-trivial potential for $\sigma$. The
competition between the constant flux term $W_0 = \langle W_{\mathrm{flux}}\rangle$
and the exponential \eqref{eq:Wnp} creates the supersymmetric AdS minimum
of the KKLT construction. Its subsequent uplift to a metastable de~Sitter
vacuum exploits the freedom in the sign of the vacuum energy that is
intrinsic to the supergravity potential \eqref{eq:sugra_potential} --- a
freedom that is absent in the globally supersymmetric theory and that was
identified in Section~\ref{sec:SUGRA} as one of the central new features introduced by
the localisation of supersymmetry.

\subsubsection*{Closure of the Arc: the Seiberg--Witten Curve Revisited}

Section~\ref{sec:strings} also closes the logical arc opened in Section~\ref{sec:SW}. The Seiberg--Witten
curve, derived in that section as an abstract holomorphic object encoding the
non-perturbative dynamics of $\mathcal{N}=2$ SU(2) Yang--Mills theory,
reappears in the string-theoretic context as the actual geometry of a brane
configuration. For the SU(2) theory realised on a pair of D3-branes separated
by a distance $2a$ in the transverse directions, the gauge-invariant modulus
$u = \mathrm{Tr}\,\varphi^2 = 2a^2$ is literally the squared separation of
the branes, and the singular loci $u = \pm\Lambda^2$ where BPS states become
massless correspond to coincident branes.

More precisely, the Seiberg--Witten differential $\lambda_{\mathrm{SW}}$,
whose period integrals
$a = \oint_\alpha \lambda_{\mathrm{SW}}$ and
$a_D = \oint_\beta \lambda_{\mathrm{SW}}$
determine the entire low-energy dynamics, can be derived from the holomorphic
three-form $\Omega_3$ of the Calabi--Yau threefold in Type~IIB compactifications:
\begin{equation}
    a = \oint_\alpha \lambda_{\mathrm{SW}},
    \qquad
    a_D = \oint_\beta \lambda_{\mathrm{SW}},
    \qquad
    \lambda_{\mathrm{SW}} = \int_{\Sigma_2} \Omega_3,
\label{eq:SW_periods_CY}
\end{equation}
where $\Sigma_2$ is a two-cycle in the internal geometry connecting the
D-brane positions. The abstract mathematical object of Section~\ref{sec:SW} is thereby
identified with a concrete geometric quantity in the string-theoretic
compactification of Section~\ref{sec:strings}.

\subsubsection*{Summary}

\begin{figure}[h]
\centering
\begin{tikzpicture}[
  node distance=1.2cm and 2.2cm,
  every node/.style={align=center},
  box/.style={
    draw, rounded corners=4pt,
    minimum width=3.8cm, minimum height=1.0cm,
    text width=3.6cm, align=center,
    font=\small
  },
  bigbox/.style={
    draw, rounded corners=4pt,
    minimum width=3.8cm, minimum height=1.6cm,
    text width=3.6cm, align=center,
    font=\small
  },
  arrow/.style={-{Stealth[length=5pt]}, thick}
]

%--- Section 3 and Section 4 (top row) ---
\node[bigbox] (sec3) {
  \textbf{Section~\ref{sec:SW}}\\
  {\footnotesize (global $\mathcal{N}=2$)}\\[4pt]
  {\footnotesize SW curve\\periods $a,\, a_D$\\prepotential $\mathcal{F}$}
};

\node[bigbox, right=of sec3] (sec4) {
  \textbf{Section~\ref{sec:SUGRA}}\\
  {\footnotesize (local $\mathcal{N}=1$)}\\[4pt]
  {\footnotesize $K,\, W,\, f_{ab}$\\scalar potential $V$\\geometry $\mathcal{M}$}
};

%--- Section 5 centered between 3 and 4 ---
\coordinate (mid34) at ($(sec3.south)!0.5!(sec4.south)$);
\node[box, below=1.4cm of mid34] (sec5) {
  \textbf{Section~\ref{sec:strings}}\\
  {\footnotesize (string theory)}
};

%--- Three branches below Section 5 ---
\node[box, below left=1.4cm and 2.0cm of sec5] (dbrane) {
  \textbf{D-branes}\\
  {\footnotesize (source of\\gauge theories)}
};

\node[box, below=1.4cm of sec5] (ads) {
  \textbf{AdS/CFT}\\
  {\footnotesize (holographic\\supergravity dual)}
};

\node[box, below right=1.4cm and 2.0cm of sec5] (cy) {
  \textbf{Calabi--Yau\\compactification}\\
  {\footnotesize (source of $K,\,W$)}
};

%--- Section 6 centered below the three branches ---
\coordinate (midbot) at ($(dbrane.south)!0.5!(cy.south)$);
\node[box, below=1.4cm of midbot] (sec6) {
  \textbf{Section~\ref{sec:moduli} }\\
  {\footnotesize (KKLT: explicit\\realization of $V$\\from Section~4,\\with $K,W$\\from Section~5)}
};

%--- Arrows: Sections 3 & 4 -> Section 5 ---
\draw[arrow] (sec3.south) -- ++(0,-0.5) -| (sec5.north west);
\draw[arrow] (sec4.south) -- ++(0,-0.5) -| (sec5.north east);

%--- Arrows: Section 5 -> branches ---
\draw[arrow] (sec5.south) -- ++(0,-0.4) -| (dbrane.north);
\draw[arrow] (sec5.south) -- (ads.north);
\draw[arrow] (sec5.south) -- ++(0,-0.4) -| (cy.north);

%--- Arrows: branches -> Section 6 ---
\draw[arrow] (dbrane.south) -- ++(0,-0.4) -| (sec6.north west);
\draw[arrow] (ads.south)    -- (sec6.north);
\draw[arrow] (cy.south)     -- ++(0,-0.4) -| (sec6.north east);

\end{tikzpicture}
\caption{Logical structure of the paper.}
\label{fig:structure}
\end{figure}
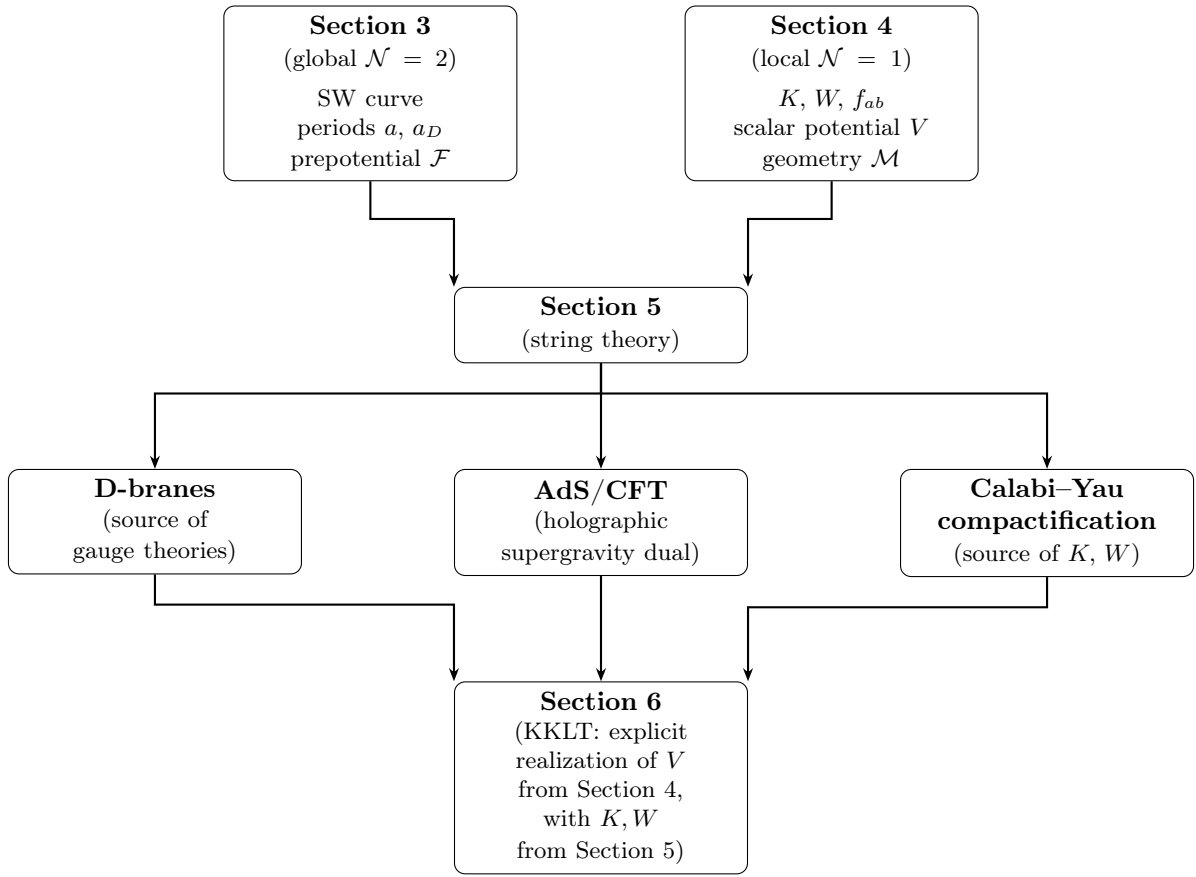

The relationship between Sections~\ref{sec:SUGRA} and~\ref{sec:strings} is that of an abstract formalism
and its geometric realisation. Section~\ref{sec:SUGRA} establishes that the entire
$\mathcal{N}=1$ supergravity Lagrangian is determined by the triple $(K, W, f_{ab})$;
Section~\ref{sec:strings} derives these functions from the geometry of the compactification
manifold $X_6$, the choice of quantised flux background, and the dynamics of
D-branes. At the same time, Section~\ref{sec:strings} completes the logical circle of the
paper by showing that the Seiberg--Witten curve of Section~\ref{sec:SW} --- derived from
purely field-theoretic considerations --- is the geometry of the brane
configuration, and that its period integrals are integrals of the holomorphic
three-form $\Omega_3$ over two-cycles in the internal space. The unifying
principle throughout is that the vacuum structure of supersymmetric theories
is encoded in the geometry of holomorphic objects, and that the physical
questions one wishes to answer correspond to geometric properties of those
objects: their singularities, their monodromy, and the sign and magnitude
of the potential they generate. The connections between Sections are presented in Figure~\ref{fig:structure}.

%%%%%%%%%%%%%%%%%%%%%%%%%%%%
\section{Moduli stabilisation and de Sitter vacua}\label{sec:moduli}
One of the central challenges in string phenomenology is the problem of moduli stabilisation. 
Compactifications of string theory to four dimensions inevitably give rise to massless scalar 
fields---moduli---parameterising the size and shape of the extra dimensions. These moduli must 
be stabilised to generate realistic low-energy physics and to avoid conflicts with observational 
constraints on fifth forces and time-variation of coupling constants.

Kachru, Kallosh, Linde, and Trivedi~\cite{KKLT2003} showed that a 
non-perturbative superpotential of the form $W = W_0 + A\,e^{-aT}$, combined 
with flux stabilisation of the complex-structure moduli, fixes the Kähler 
modulus at a supersymmetric AdS minimum; a subsequent anti-D3-brane 
contribution uplifts this to a metastable de~Sitter vacuum. However, the KKLT scenario relies on several delicate ingredients whose consistency remains 
under active investigation:

\begin{enumerate}
    \item \textbf{Control of approximations:} The construction requires working in a regime 
    where both $\alpha'$ and $g_s$ corrections are under control, necessitating large volume 
    and weak coupling.
    
    \item \textbf{$\alpha'$ corrections:} Higher-derivative corrections to the K\"ahler potential, 
    particularly the $\alpha'^3$ correction computed by Becker et al.~\cite{Becker:2002nn}, 
    modify the no-scale structure and can significantly affect the vacuum structure.
    
    \item \textbf{The swampland programme:} Recent conjectures about the landscape of consistent 
    quantum gravity theories, particularly the de Sitter swampland conjecture~\cite{Obied:2018sgi, 
    Ooguri:2018wrx}, pose a fundamental challenge to the existence of metastable de Sitter vacua 
    in string theory.
\end{enumerate}

The tension between these requirements is acute. On one hand, control of the effective field 
theory demands large volume $\mathcal{V} \gg 1$ (in string units) and weak coupling $g_s \ll 1$. 
On the other hand, the de Sitter swampland conjecture suggests that such controlled de Sitter 
constructions may be impossible in a consistent theory of quantum gravity.

We aim to clarify these issues by:

\begin{itemize}
    \item Systematically analysing the effect of $\alpha'^3$ corrections to the K\"ahler potential on the KKLT vacuum structure;
    
    \item Quantifying the tension between the KKLT construction and the de Sitter swampland 
    conjecture;
    
    \item Mapping out the parameter space where metastable de Sitter minima exist versus where the theory enters the swampland;
    
    \item Identifying the critical values of correction parameters where the vacuum structure 
    undergoes qualitative changes.
\end{itemize}

Understanding whether controlled de Sitter vacua exist in string theory is not merely a 
technical question---it bears directly on the viability of string theory as a framework for 
describing our accelerating universe. If the swampland conjectures are correct, they would 
rule out not only KKLT but also large classes of inflationary models and dark energy scenarios. 
Conversely, if explicit counterexamples can be constructed with full control over all corrections, 
this would constrain the formulation of swampland criteria.

The present analysis focuses on the interplay between perturbative $\alpha'$
corrections and non-perturbative effects in the simplest setting: single-modulus
compactifications. Despite its simplicity, this framework captures the essential
physics and allows for analytic control over the minimisation conditions. The
results generalise straightforwardly to multi-modulus scenarios and provide a
foundation for understanding more sophisticated constructions such as the Large
Volume Scenario~\cite{Balasubramanian:2005zx}.

\subsection{The KKLT Construction}
General Kahler potential in Type IIB
% Общий кэлеров потенциал в Type IIB
\begin{equation}
K = -\ln\left(-i\int_{X_6} \Omega \wedge \bar{\Omega}\right) - 2\ln\mathcal{V} + K_{\text{matter}}
\end{equation}
The volume of the Calabi--Yau is $\mathcal{V} = (2\sigma)^{3/2}/4$.
After stabilization of complex structures
% После стабилизации комплексных структур
\begin{equation}
K = \langle K_{\text{cs}} \rangle - 3\ln(2\sigma) + K_{\text{matter}}
\end{equation}
In the simplest KKLT (one module, without material fields)
% В простейшем KKLT (один модуль, без материальных полей)
\begin{equation}
K_{\text{cl}} = -3\ln(2\sigma)
\end{equation}

One of the central challenges in connecting string theory to phenomenology is the stabilization of moduli---the massless scalar fields arising from the compactification of extra dimensions. In the context of Type IIB string theory compactified on Calabi--Yau orientifolds, the flux compactification framework provides a mechanism to stabilize the complex structure moduli and the dilaton, leaving the K\"ahler moduli (specifically the volume modulus) massless at the classical level.

The seminal work of Kachru, Kallosh, Linde, and Trivedi (KKLT) \cite{KKLT2003}  proposed a mechanism to stabilize the K\"ahler modulus $T = \sigma + i\theta$ using non-perturbative effects. The superpotential takes the form:
\begin{equation}
    W = W_0 + A e^{-aT},
\end{equation}
where $W_0$ is the constant flux superpotential and the exponential term arises from gaugino condensation or Euclidean D3-brane instantons.

However, the classical K\"ahler potential $K_{\text{cl}} = -3\ln(T + \bar{T})$ exhibits a "no-scale" structure, characterized by the identity 
\begin{equation}
K^{T\bar{T}}(\partial_T K)(\partial_{\bar{T}} K) = 3. 
\label{eq:no_scale}
\end{equation}
This symmetry ensures that the classical scalar potential vanishes exactly at any supersymmetric point ($D_T W = 0$), meaning the volume modulus remains a flat direction and is not stabilized by the classical theory alone. The potential is identically zero at the classical level. Consequently, quantum correc-
tions are indispensable for moduli stabilisation: without them, the volume modulus
has no potential at all. Classically, this means that the modulus $T$ is a flat direction; it has no potential at all—it can take any value, and the theory is scale invariant at the classical level.

To break this no-scale structure and generate a potential for the volume modulus, quantum corrections are indispensable. 
The condition $D_T W = 0$ with $\partial_T W = -aAe^{-aT}$ gives,
in the large-volume regime $a\sigma_0 \gg 1$,
\begin{equation}\label{eq:KKLT_min}
  W_0 \;\approx\; -A\,e^{-a\sigma_0}\!
  \left(1 + \frac{2a\sigma_0}{3}\right).
\end{equation}
This relation determines $\sigma_0$ implicitly through the
hierarchy $|W_0/A|$. The vacuum energy at this critical point is
\begin{equation}\label{eq:V_AdS}
  \V_{AdS}
  \;=\; -3\,e^{K_0}|W_0|^2
  \;<\; 0,
  \qquad
  e^{K_0} \;=\; \frac{1}{8\sigma_0^3},
\end{equation}
which establishes a supersymmetric anti-de~Sitter minimum. To obtain a cosmological vacuum with a positive cosmological constant, this AdS vacuum must be uplifted to a metastable de Sitter state, typically by adding anti-D3-branes or other uplifting mechanisms.

As we have seen, in the classical theory (without corrections), the Kähler potential has a special form $K_{\text{cl}} = -3\ln(2\sigma)$. A correction arising from 10-dimensional string theory
of order $\alpha'^3$~\cite{Becker:2002nn}:
\begin{equation}\label{eq:10d_R4}
  S_{\alpha'^3} \;\supset\;
  -\frac{(\alpha')^3}{3\cdot 2^{10}}
  \int d^{10}x\,\sqrt{-g^{(10)}}\,e^{-2\phi}\,\mathcal{R}^4,
\end{equation}
where $\mathcal{R}^4 = t_8 t_8 R^4 +
\tfrac{1}{8}\varepsilon_{10}\varepsilon_{10}R^4$ denotes a specific contraction of four Riemann curvature tensors \cite{Becker:2002nn}, 
yields a correction to the four-dimensional K\"{a}hler potential~\cite{Becker:2002nn}:

\begin{equation}\label{eq:dK}
  \delta K^{(\alpha'^3)} \;=\; -\,\frac{\hatxi}{2\cV},
  \qquad
  \hatxi \;\equiv\; -\,\frac{\chi(X_6)}{2(2\pi)^3}
  \cdot\frac{\zeta(3)}{g_s^{3/2}},
\end{equation}
\noindent
where $\chi(X_6)$ is the Euler characteristic of the Calabi--Yau,
$\zeta(3) \approx 1.202$ is Ap\'{e}ry's constant, and $g_s$ is the
string coupling.

\begin{remark}\label{rem:sign}
  The sign of $\hatxi$ is determined by the sign of $\chi(X_6)$.
  For typical Calabi--Yau manifolds with $h^{2,1} > h^{1,1}$ one
  has $\chi < 0$, hence $\hatxi > 0$. This is the physically
  relevant regime and will be assumed throughout.
\end{remark}

Including the $\alpha'^3$ correction, the full K\"{a}hler potential
for the single-modulus case becomes
\begin{equation}\label{eq:K_full}
  K \;=\; K_{\mathrm{cl}} + \delta K^{(\alpha'^3)}
  \;=\; -3\ln(2\sigma) \;-\; \frac{\hatxi}{16\sigma^{3/2}},
\end{equation}
where the last term uses $\cV = (2\sigma)^{3/2}/4$.
\subsection*{Self-consistency of the uplift}

After adding the anti-D3-brane contribution
\begin{equation}
    V_{\rm total}(\sigma) = V(\sigma) + \frac{D}{\sigma^3}, \qquad D > 0,
    \label{eq:uplift_total}
\end{equation}
the minimum shifts from $\sigma_0$ to a new value $\sigma_1$. To first order in $D$, expanding
$\partial V_{\rm total}/\partial\sigma = 0$ around $\sigma_0$ gives
\begin{equation}
    \frac{\delta\sigma}{\sigma_0} \equiv \frac{\sigma_1 - \sigma_0}{\sigma_0}
    \approx \frac{3D/\sigma_0^4}{V''(\sigma_0)},
    \label{eq:shift_uplift}
\end{equation}
where $V''(\sigma_0) \equiv \partial^2 V/\partial\sigma^2\big|_{\sigma_0} > 0$ is the modulus
mass squared at the pre-uplift minimum.
The uplift is self-consistent if and only if
\begin{equation}
    \frac{\delta\sigma}{\sigma_0} \ll 1
    \quad\Longleftrightarrow\quad
    D \ll \tfrac{1}{3}\,\sigma_0^4\,V''(\sigma_0).
    \label{eq:selfconsistency}
\end{equation}
The uplift parameter $D$ is fixed by requiring a Minkowski (or de~Sitter) vacuum,
$V_{\rm total}(\sigma_1) \approx 0$, which at leading order gives
\begin{equation}
    D \approx |V_{\rm AdS}(\sigma_0)|\,\sigma_0^3
      = \frac{3\,|W_0|^2}{8\,\sigma_0^3}\cdot\sigma_0^3
      = \frac{3\,|W_0|^2}{8}.
    \label{eq:D_uplift}
\end{equation}
The modulus mass at the AdS minimum is
\begin{equation}
    V''(\sigma_0) \sim \frac{a^2\,|W_0|^2}{\sigma_0^3},
    \label{eq:modmass}
\end{equation}
so the self-consistency condition~\eqref{eq:selfconsistency} becomes
\begin{equation}
    \frac{\delta\sigma}{\sigma_0}
    \sim \frac{1}{a^2\sigma_0} \ll 1,
    \label{eq:consistency_final}
\end{equation}
which is satisfied whenever $a\sigma_0 \gg 1$.
This is precisely the large-volume regime in which the KKLT construction is valid, so the
uplift does not destabilise the modulus: the shift $\delta\sigma/\sigma_0 \sim (a\sigma_0)^{-1}$
is suppressed by the same hierarchy that justifies the non-perturbative approximation.
\subsection{Scalar Potential Structure}

The correction $\delta K^{(\alpha'^3)}$ breaks the no-scale identity~\eqref{eq:no_scale}:
\begin{equation}\label{eq:no_scale_broken}
  K^{T\bar{T}}(\partial_T K)(\partial_{\bar{T}} K)
  \;=\; 3 \;-\; \frac{9\hatxi}{8\cV}
  \;+\; \mathcal{O}\!\left(\frac{\hatxi^2}{\cV^2}\right).
\end{equation}
The deviation from~$3$ is suppressed by $\cV^{-1}$ and hence
small when $\cV \gg \hatxi$. Nevertheless it is conceptually
crucial: it activates the $-3|W|^2$ term in the scalar potential
that previously cancelled exactly against the kinetic term, thereby
generating a non-trivial scalar potential for $\sigma$ even before
non-perturbative effects are included.

Substituting $K$~\eqref{eq:K_full} and $W$~\eqref{eq:W_KKLT} into
$V = e^{K/\Mpl^2}\!\left(K^{T\bar{T}}|D_T W|^2 - 3|W|^2/\Mpl^2\right)$
(the KW-form of eq.~\eqref{eq:scalar_potential_alt}), evaluating
$D_T W = \partial_T W + (\partial_T K)W$, and minimising over the
axion $\theta$ (which gives $\cos(a\theta_0 + \phi) = +1$,
$\phi = \arg(A/W_0)$), we obtain the following explicit formula:

\begin{equation}\label{eq:V_sigma}
  V(\sigma) \;=\;
  \underbrace{\frac{a^2|A|^2\,e^{-2a\sigma}}{2\sigma}}_{V^{(1)}_{\mathrm{np}}}
  \;-\;
  \underbrace{\frac{a|A||W_0|\,e^{-a\sigma}}{2\sigma^2}}_{V^{(2)}_{\mathrm{np}}}
  \;+\;
  \underbrace{\frac{3\hatxi\,|W_0|^2}{8\sigma^3}}_{\displaystyle V_\xi\;[\text{new}]}.
\end{equation}

\begin{remark}\label{rem:newterm}
  The first two terms in~\eqref{eq:V_sigma} are present in
  classical \KKLT{} ($\hatxi = 0$). The third term
  $V_\xi \propto \hatxi\,\sigma^{-3}$ is entirely new.
  For $\hatxi > 0$ it is \emph{positive} and competes with
  the negative contribution $-V^{(2)}_{\mathrm{np}}$,
  thereby shifting the location and depth of the AdS minimum.
\end{remark}

The minimum exists as long as the Hessian condition $\partial^2 V/\partial\sigma^2\big|_{\sigma_0} > 0$ is satisfied. A
detailed analysis of the two constraints $\partial V/\partial\sigma = 0$ and $\partial^2 V/\partial\sigma^2 > 0$ yields a critical
threshold
\begin{equation}
    \hat{\xi}_c \approx \frac{(a\sigma_0^{\rm cl})^{3/2}}{3\sqrt{2}}\,
    \frac{|A|\,e^{-a\sigma_0^{\rm cl}}}{|W_0|}.
    \label{eq:14}
\end{equation}
Since in the KKLT regime $|W_0| \ll |A|e^{-a\sigma_0^{\rm cl}}$, one has $\hat{\xi}_c \gg 1$ (in units of $M_{\rm Pl}$). Thus,
for typical KKLT parameters the AdS minimum \textit{survives} the $\alpha'^3$ correction, merely
shifting to a larger volume.

\subsection{Three Regimes of the Potential}
\label{sec:three_regimes}

The potential~\eqref{eq:V_sigma} depends on a single dimensionless
parameter $\hatxi$, which controls the strength of the $\alpha'^3$
correction relative to the non-perturbative terms. As $\hatxi$ increases
from zero, the structure of the potential changes qualitatively at the
critical value~\eqref{eq:14}, defining three physically distinct regimes
summarised in Table~\ref{tab:regimes}.

\begin{table}[ht]
\centering
\caption{Three regimes of the \KKLT{} scalar potential~\eqref{eq:V_sigma}
as a function of the $\alpha'^3$ correction parameter $\hatxi$.
The critical threshold $\hatxi_c$ is given by~\eqref{eq:14}.}
\renewcommand{\arraystretch}{1.3}
\begin{tabular}{clll}
\hline
Regime & Condition & Vacuum & Physics \\
\hline
I   & $\hatxi = 0$
    & AdS minimum at $\sigma_0^{\mathrm{cl}}$
    & classical \KKLT{} \\
II  & $0 < \hatxi < \hatxi_c$
    & AdS minimum at $\sigma_0 > \sigma_0^{\mathrm{cl}}$
    & \KKLT{} with $\alpha'^3$ correction \\
III & $\hatxi > \hatxi_c$
    & no minimum (runaway)
    & LVS or decompactification \\
\hline
\end{tabular}
\label{tab:regimes}
\end{table}

\paragraph{Regime~I: Classical \KKLT{} ($\hatxi = 0$).}

When $\alpha'$ corrections are absent the potential reduces to the two
non-perturbative terms $V^{(1)}_{\mathrm{np}} - V^{(2)}_{\mathrm{np}}$
of~\eqref{eq:V_sigma}. Their competition creates a unique minimum at
$\sigma_0^{\mathrm{cl}}$ determined implicitly by~\eqref{eq:KKLT_min}.
At this minimum the Kähler-covariant derivative vanishes,
$D_T W\big|_{\sigma_0^{\mathrm{cl}}}=0$, so supersymmetry is
preserved and the vacuum energy is
\begin{equation}
  V_{\mathrm{AdS}}^{\mathrm{cl}}
  = -3\,e^{K_0}|W_0|^2
  = -\frac{3|W_0|^2}{8(\sigma_0^{\mathrm{cl}})^3} < 0,
\end{equation}
giving a supersymmetric Anti-de~Sitter vacuum. The gravitino mass is
\begin{equation}\label{eq:m32_I}
  m_{3/2}
  = e^{K_0/2}|W_0|\,M_{\mathrm{Pl}}
  = \frac{|W_0|}{2\sqrt{2}\,(\sigma_0^{\mathrm{cl}})^{3/2}}\,M_{\mathrm{Pl}},
\end{equation}
which is hierarchically small when $|W_0|\ll 1$, the key phenomenological
virtue of the construction. Uplifting to de~Sitter requires the addition of
an $\overline{\mathrm{D3}}$-brane term $D/\sigma^3$ with $D\approx
|V_{\mathrm{AdS}}^{\mathrm{cl}}|\,(\sigma_0^{\mathrm{cl}})^3$, shifting
the vacuum energy to a small positive value while leaving the modulus
stabilised at approximately the same $\sigma_0^{\mathrm{cl}}$.

\paragraph{Regime~II: Corrected \KKLT{} ($0<\hatxi<\hatxi_c$).}

The new term $V_\xi = 3\hatxi|W_0|^2/(8\sigma^3)$ is positive and grows
as $\sigma$ decreases, pushing the minimum towards larger volume. The
corrected minimum $\sigma_0$ satisfies $\sigma_0 > \sigma_0^{\mathrm{cl}}$
by an amount that grows with $\hatxi$. Since the non-perturbative terms
fall exponentially at large $\sigma$ while $V_\xi\propto\sigma^{-3}$
decays only as a power law, the net effect is a shallower AdS minimum,
\begin{equation}\label{eq:VAdS_II}
  |V_{\mathrm{AdS}}(\sigma_0)|
  < |V_{\mathrm{AdS}}^{\mathrm{cl}}|,
\end{equation}
and a lighter modulus mass $m_T^2 = \partial^2 V/\partial\sigma^2|_{\sigma_0}$.
The gravitino mass~\eqref{eq:m32_I} is reduced accordingly, since
$e^{K_0/2}\propto\sigma_0^{-3/2}$ and $\sigma_0$ has increased. Both the
qualitative structure and the phenomenological hierarchy
$m_{3/2}\ll M_{\mathrm{Pl}}$ are preserved throughout this regime, so the
$\alpha'^3$ correction modifies the quantitative predictions of \KKLT{}
without invalidating the mechanism. The uplift to de~Sitter proceeds as
before, now with a corrected value of $D$.

\paragraph{Regime~III: Runaway ($\hatxi>\hatxi_c$).}

Once $\hatxi$ exceeds the threshold~\eqref{eq:14}, the positive
contribution $V_\xi$ overwhelms the non-perturbative attraction at all
$\sigma$, and the Hessian condition $\partial^2 V/\partial\sigma^2>0$
can no longer be satisfied. The potential develops a runaway towards
large volume, $V\to 0^-$ as $\sigma\to\infty$. This signals either a
transition to the Large Volume
Scenario~\cite{Balasubramanian:2005zx}, where the minimum is
stabilised by the interplay of $V_\xi$ with subleading perturbative
corrections at exponentially large volume, or outright decompactification
if no such stabilisation occurs.

The boundary $\hatxi=\hatxi_c$ is therefore a phase boundary in the
space of string compactifications: on one side lies a controlled \KKLT{}
vacuum, on the other a runaway. Since
$\hatxi\propto -\chi(X_6)/g_s^{3/2}$ grows as the string coupling
decreases or as one moves to Calabi--Yau manifolds with larger
$|h^{2,1}-h^{1,1}|$, this phase boundary is a concrete constraint on
the class of compactifications that admit metastable de~Sitter vacua
within the \KKLT{} framework. Confronting it with the de~Sitter
swampland conjecture~\cite{Obied:2018sgi,Ooguri:2018wrx}, which
asserts $|\nabla V|\geq c\,V$ for some order-one constant $c$, one
finds that the swampland bound is most dangerous near the boundary:
precisely when $\hatxi\lesssim\hatxi_c$ the potential becomes very
flat near its minimum, making it harder to satisfy the gradient
condition for any positive uplift. This tension --- between the need
for a shallow minimum to obtain a small positive cosmological constant
and the swampland requirement of a sufficiently steep gradient ---
represents one of the central unresolved questions in string
cosmology, and the three-regime structure of the potential mapped out
here provides a concrete arena in which to study it.

%%%%%%%%%%%%%%%%%%%%%%%%%%%%%%%%%%
\section{Conclusion}\label{sec:conclusion}

The thread running through this work is the idea that supersymmetry, far from
being merely a device for improving ultraviolet behaviour, provides a window into
the exact non-perturbative structure of quantum field theory. We have followed this
thread from the algebra itself --- the anticommutator
$\{Q_\alpha,\bar{Q}_{\dot\alpha}\}=2\sigma^\mu_{\alpha\dot\alpha}P_\mu$ that
links fermionic symmetries to spacetime translations --- through its realisation
in $\mathcal{N}=1$ Yang--Mills and the Wess--Zumino model, into the richer
territory of $\mathcal{N}=2$ dynamics, and finally up to the string-theoretic
and cosmological questions surrounding moduli stabilisation and de~Sitter vacua.

The centrepiece of the analysis is the Seiberg--Witten
solution~\cite{Seiberg1994a,Seiberg1994b}. What makes it remarkable is not merely
that it produces exact results, but that it translates a dynamical question ---
what is the low-energy physics of $\mathcal{N}=2$ $SU(2)$ Yang--Mills at strong
coupling? --- into a geometric one: what are the period integrals
\begin{equation}\label{eq:periods_concl}
  a = \oint_{\alpha}\lambda_{\mathrm{SW}}, \qquad
  a_D = \oint_{\beta}\lambda_{\mathrm{SW}}
\end{equation}
of the meromorphic differential $\lambda_{\mathrm{SW}}=x\,dx/y$ over the two
cycles of the elliptic curve $\Sigma_u$? The prepotential $\mathcal{F}(a)$, the
BPS mass formula $M=|n_e a + n_m a_D|$, and the electromagnetic duality group
$SL(2,\mathbb{Z})$ all follow from this single geometric object. The Picard--Fuchs
equation of Section~\ref{sec:SW} organises the analytic continuation of the
periods around the singular loci $u=\pm\Lambda^2$ and $u=\infty$ into a precise
set of monodromy matrices, making manifest that electromagnetic duality is not an
assumption but a consequence of the topology of the moduli space. The proof of
confinement in this framework --- a soft mass $\mu\,\mathrm{Tr}\,\varphi^2$
selects vacua where a BPS monopole condenses, producing the dual Meissner effect
and confining electric flux into tubes --- remains, three decades later, the only
exact and fully controlled demonstration of this mechanism in four-dimensional
gauge theory.

The transition to supergravity, worked out in Section~\ref{sec:SUGRA}, shows that
the two holomorphic functions $K$ and $W$ play a role analogous to the prepotential
in the $\mathcal{N}=2$ theory: once they are specified, every sector of the
Lagrangian is fixed. The five sectors --- kinetic terms, gauge dynamics encoded
in $f_{ab}(\Phi)$, the scalar potential, Yukawa couplings, and the four-fermion
contact interaction proportional to the Riemann curvature of the scalar manifold
--- are all determined by $K$, $W$, and $f_{ab}$ through the superspace integrals
of the master action. The key new feature relative to the global theory is the
scalar potential
\begin{equation}\label{eq:V_concl}
  V = M_{\mathrm{Pl}}^4\,e^{G}\!\left(g^{i\bar{\jmath}}G_i G_{\bar{\jmath}}
      - 3\right), \qquad
  G = \frac{K}{M_{\mathrm{Pl}}^2} + \ln\frac{|W|^2}{M_{\mathrm{Pl}}^6},
\end{equation}
where the exponential factor $e^{K/M_{\mathrm{Pl}}^2}$ and the negative
gravitational term $-3|W|^2/M_{\mathrm{Pl}}^2$ together allow the vacuum energy
to take any sign. This is not a complication to be overcome but a feature: it is
precisely this freedom that makes the uplift from a supersymmetric AdS minimum to
a de~Sitter vacuum possible.

Section~\ref{sec:strings} placed both constructions --- Seiberg--Witten theory and
$\mathcal{N}=1$ supergravity --- in the context of string theory. D3-branes in
flat space carry $\mathcal{N}=4$ Yang--Mills on their worldvolume; the Seiberg--Witten
curve re-emerges as the geometry of the brane configuration when the branes are
separated in the transverse directions, and the BPS mass formula is reproduced by
the tension of fundamental strings and D1-branes stretched between them. The
AdS/CFT correspondence then identifies the strong-coupling limit of this gauge
theory with classical supergravity on $\mathrm{AdS}_5\times S^5$, providing a
complementary, holographic window into the same non-perturbative physics. The
reduction from $\mathcal{N}=4$ to phenomenologically viable $\mathcal{N}=1$ models
can be achieved by three structurally distinct mechanisms --- orbifold projections
of the internal geometry, explicit mass deformations of the adjoint chiral
multiplets, and the threading of quantised $G_3$ fluxes through the cycles of the
Calabi--Yau. In each case the K\"{a}hler potential and superpotential of the
effective four-dimensional theory descend from the geometry of the compactification
manifold, giving a concrete geometric realisation of the abstract field-theoretic
structures of Section~\ref{sec:SUGRA}.

The analysis of Section~6 addressed one of the central open problems that this
string-theoretic picture raises. Compactifications generically produce massless
scalar moduli that must be stabilised to yield a consistent low-energy physics.
The KKLT construction~\cite{KKLT2003} achieves this by combining a flux-generated
constant $W_0$ with a non-perturbative term $Ae^{-aT}$ to produce a supersymmetric
AdS minimum, which is then uplifted to a metastable de~Sitter vacuum by the
positive energy of $\overline{\mathrm{D3}}$-branes. We showed that the
$\alpha'^3$ correction~\cite{Becker:2002nn} to the K\"{a}hler potential, which
shifts the no-scale identity $K^{T\bar{T}}(\partial_T K)(\partial_{\bar{T}}K)$
away from~$3$ by a term of order $\hatxi/\mathcal{V}$, does not destroy the AdS
minimum for typical KKLT parameters: it merely displaces it to a somewhat larger
volume and modifies the depth of the potential. The critical threshold
$\hatxi_c \approx (a\sigma_0)^{3/2}|A|e^{-a\sigma_0}/(3\sqrt{2}|W_0|)$
determines when the correction becomes large enough to remove the minimum
entirely, in which case the potential develops a runaway towards large volume ---
the regime of the Large Volume Scenario~\cite{Balasubramanian:2005zx} --- or
towards complete decompactification. The three regimes of Table~\ref{tab:regimes}
thus map out the parameter space of KKLT compactifications under perturbative
$\alpha'$ corrections and provide a concrete diagnostic for distinguishing
controlled de~Sitter constructions from configurations that belong to the
swampland~\cite{Obied:2018sgi,Ooguri:2018wrx}.

Several questions raised by this work remain open. On the gauge theory side, the
extension of the Seiberg--Witten solution to higher-rank gauge groups and to
theories with matter hypermultiplets involves a significant enrichment of the
geometry: the elliptic curve is replaced by a higher-genus Riemann surface, and
the periods become integrals of an abelian differential over a surface of genus
$r = \mathrm{rank}(G)$. The systematic computation of instanton corrections to
the prepotential via Nekrasov's localisation formula, and the AGT correspondence
that relates $\mathcal{N}=2$ partition functions on $\mathbb{R}^4$ to
Liouville correlation functions on a punctured torus, remain active areas where
the geometric and algebraic structures of the present review continue to generate
new results. On the supergravity and string theory side, the conditions under
which flux compactifications yield de~Sitter vacua with full control over the
$\alpha'$ and $g_s$ expansion are not yet settled, and the precise status of the
de~Sitter swampland conjecture --- whether it rules out metastable de~Sitter
vacua or merely constrains their parametric regime --- continues to be debated.
These are not peripheral questions. They bear directly on whether superstring
theory can account for the observed accelerating expansion of the universe, and
on the extent to which the landscape of string vacua is constrained by
consistency requirements of quantum gravity.

What the present work has tried to make clear is that these questions are not
isolated: they are facets of a single geometric programme that began with the
Seiberg--Witten solution and has since expanded to encompass supergravity,
string compactifications, and cosmology. The unifying principle is that the
vacuum structure of supersymmetric theories is encoded in the geometry of
holomorphic objects --- curves, prepotentials, K\"{a}hler potentials,
superpotentials --- and that the physical questions one wishes to answer
correspond to geometric properties of these objects: their singularities, their
monodromy, the sign and magnitude of the potential they generate. This
translation of physics into geometry, and back, is what gives the subject its
distinctive character and its continuing vitality.

%%%%%%%%%%%%%%%%%%%%%%%%%

\phantomsection
\section*{Outlook}
\addcontentsline{toc}{section}{Outlook}

The analysis carried out in this work opens several concrete directions for future investigation,
spanning all levels of the hierarchy from gauge dynamics to string cosmology.

\subsection*{Non-perturbative gauge dynamics (Section~\ref{sec:SW})}

The non-perturbative solution of Section~\ref{sec:SW} was developed for the simplest case of
$\mathcal{N}=2$ $\mathrm{SU}(2)$ Yang--Mills theory without matter hypermultiplets.
Several natural generalisations remain to be explored.

\begin{itemize}

\item \textbf{Higher-rank gauge groups.}
For gauge group $G$ of rank $r$, the Seiberg--Witten curve is a hyperelliptic curve of
genus $r$, and the period integrals $a_i$, $a_{D,i}$ ($i=1,\ldots,r$) are integrals of
an abelian differential over a Riemann surface of genus $r$.  The moduli space of vacua
becomes an $r$-dimensional special Kähler manifold, and the monodromy group is a subgroup
of $\mathrm{Sp}(2r,\mathbb{Z})$.  A systematic study of the BPS spectrum, wall-crossing
phenomena, and confinement mechanisms for $G = \mathrm{SU}(N)$, $\mathrm{SO}(N)$, and
$\mathrm{Sp}(N)$ would extend the results of Section~3 to phenomenologically relevant
gauge groups.

\item \textbf{Theories with matter hypermultiplets.}
The addition of $N_f$ fundamental hypermultiplets introduces branch cuts in the
prepotential and modifies the one-loop running.  For $N_f < 4$ the theory is asymptotically
free; at $N_f = 4$ it is conformal and the Seiberg--Witten curve acquires an additional
$\mathrm{SL}(2,\mathbb{Z})$ duality acting on the coupling.  A detailed analysis of the
interplay between electric--magnetic duality and flavour symmetry in this setting
would complement the pure gauge case treated here.

\item \textbf{Nekrasov partition functions and the AGT correspondence.}
The exact instanton coefficients $\mathcal{F}_k$ in the prepotential expansion~\ref{eq:poten} can be
computed systematically via Nekrasov's $\Omega$-background regularisation.  The
Alday--Gaiotto--Tachikawa (AGT) correspondence then identifies the $\mathcal{N}=2$
partition function on $\mathbb{R}^4_\epsilon$ with a Liouville correlation function on
a punctured torus, providing a bridge between gauge theory, integrable systems, and
two-dimensional conformal field theory.  Exploring this correspondence in the presence of
surface operators and defects would further illuminate the geometric structures of
Section~3.

\end{itemize}

\subsection*{Supergravity on curved scalar manifolds (Section~\ref{sec:SUGRA})}

The supergravity formalism of Section~\ref{sec:SUGRA} treats $K$, $W$, and $f_{ab}$ as abstract inputs.
Several aspects of this framework deserve deeper analysis.

\begin{itemize}

\item \textbf{Non-perturbative corrections to $K$ and $W$.}
In the globally supersymmetric theory, holomorphy and the non-renormalisation theorems
protect $W$ from perturbative corrections beyond one loop.  Upon coupling to supergravity,
gravitational loops and threshold corrections to the Kähler potential can modify $K$ in
ways that are not protected.  A systematic classification of the corrections consistent
with the symmetries of $\mathcal{N}=1$ supergravity—in particular those arising in
heterotic and Type~II compactifications—would sharpen the connection between
Section~\ref{sec:SUGRA} and Section~\ref{sec:strings}.

\item \textbf{Kähler curvature and four-fermion scattering.}
The four-fermion contact term $\mathcal{L}_{4f} = -R_{i\bar\jmath k\bar l}\,\psi^i\psi^k
\bar\psi^{\bar\jmath}\bar\psi^{\bar l}$ of equation~\ref{eq:sector5} is a direct geometric probe of
the curvature of the scalar manifold $\mathcal{M}$.  Computing this term explicitly for
the moduli spaces arising in flux compactifications—where $\mathcal{M}$ is a product of
special Kähler and quaternionic-Kähler manifolds—and comparing with amplitudes computed
via the AdS/CFT correspondence would provide a concrete test of the structures developed
in Section~\ref{sec:SUGRA}.

\end{itemize}

\subsection*{String-Theoretic Realisations (Section~\ref{sec:strings})}

Section~\ref{sec:strings} showed how D-brane configurations and Calabi--Yau compactifications give a
geometric origin to the field-theoretic data.  Several open problems arise at this level.

\begin{itemize}

\item \textbf{Geometric engineering beyond $\mathrm{SU}(2)$.}
The identification of the Seiberg--Witten curve with the actual geometry of a brane
configuration, established for $\mathrm{SU}(2)$ in Section~5.3, should be extended to
higher-rank theories.  In the Type~IIA description, the relevant objects are M5-branes
wrapping Riemann surfaces in the M-theory lift, and the Seiberg--Witten differential
$\lambda_{\mathrm{SW}}$ appears as the pullback of the M-theory three-form.  Working out
this correspondence systematically for $\mathrm{SU}(N)$ theories with matter, including
the contribution of D6-branes as flavour sources, is a natural continuation of the
material in Section~5.3.

\item \textbf{Holographic description of confinement.}
The confinement mechanism of Section~3.6—monopole condensation triggered by a soft
mass $\mu\,\mathrm{Tr}\,\varphi^2$—has a holographic counterpart in the AdS/CFT
framework: the deformation of AdS$_5\times S^5$ to a geometry with a confining IR
wall.  Making this correspondence precise, and in particular identifying the holographic
dual of the BPS monopole that condenses, would connect the exact field-theoretic result
of Section~3 with the supergravity framework of Section~5.2.

\item\textbf{From $\mathcal{N}=4$ to realistic models.}
The three reduction mechanisms discussed in Section~5.4—orbifold projections, mass
deformations, and flux compactifications—each produce $\mathcal{N}=1$ effective theories
with specific gauge groups and matter content.  A systematic classification of the
resulting quiver gauge theories, together with an analysis of their moduli spaces and
superpotentials, would provide a catalogue of string-derived models suitable for
phenomenological applications to the MSSM and beyond.

\end{itemize}

\subsection*{Moduli Stabilisation and the Swampland
(Section~\ref{sec:moduli})}

Section~\ref{sec:moduli} identified three regimes of the KKLT scalar potential under $\alpha'^3$
corrections and determined the critical threshold $\hat\xi_c$ separating controlled
de~Sitter vacua from runaway behaviour.  The following directions build directly on
this analysis and connect it to the rest of the paper.

\begin{itemize}

\item \textbf{Multi-modulus generalisations.}
The single-modulus analysis of Section~\ref{sec:moduli} captures the essential competition between
non-perturbative and perturbative contributions to the scalar potential, but realistic
Calabi--Yau compactifications have $h^{1,1} > 1$ Kähler moduli.  Extending the
three-regime structure of Table~\ref{tab:regimes} to the multi-modulus case requires analysing the
full Hessian of the potential in the moduli space and identifying the analogue of
$\hat\xi_c$ as a condition on the matrix of $\alpha'^3$ coefficients.  This is the
natural setting for the Large Volume Scenario~\cite{Balasubramanian:2005zx}, which
arises precisely in the Regime~III runaway of the single-modulus case.

\item \textbf{Higher-order $\alpha'$ and $g_s$ corrections.}
The $\alpha'^3$ correction of Becker et al.\ is the leading perturbative correction to
the Kähler potential, but it is not the only one.  Corrections at order $\alpha'^4$ and
beyond, as well as $g_s$ loop corrections to both $K$ and $W$, modify the location
and depth of the AdS minimum computed in Section~6.1.  A systematic expansion in
$\hat\xi/\mathcal{V}$ and $g_s$ would determine the radius of convergence of the KKLT
approximation and set the boundary of the regime in which the construction is under
perturbative control.

\item \textbf{Uplift mechanisms beyond anti-D3-branes.}
The uplift to a metastable de~Sitter vacuum in Section~5.4 and Section~6.1 relies on
the positive energy contribution $D/\sigma^3$ of anti-D3-branes.  Alternative uplift
mechanisms—$F$-term uplifts via nilpotent chiral superfields, $D$-term uplifts via
anomalous $\mathrm{U}(1)$ gauge fields, and $T$-brane configurations—each modify the
functional form of $V_{\mathrm{total}}$ and therefore shift the phase boundary at
$\hat\xi_c$.  Comparing the three-regime structure of Table~\ref{tab:regimes} across these different
uplift mechanisms would provide a more complete map of the KKLT parameter space.

\item \textbf{Quantitative confrontation with the swampland conjectures.}
Section~6.3 noted qualitatively that the de~Sitter swampland conjecture
$|\nabla V|\geq c\,V$ is most dangerous near the phase boundary $\hat\xi\lesssim\hat\xi_c$,
where the potential flattens.  A quantitative analysis should evaluate the ratio
$|\partial_\sigma V|/V$ at the uplifted de~Sitter minimum as a function of $\hat\xi$
and compare it with the conjectured lower bound $c\sim\mathcal{O}(1)$.  Combined with
the refined de~Sitter conjecture~\cite{Ooguri:2018wrx}, which also constrains the
Hessian eigenvalues, this would determine precisely which region of the $(\hat\xi,\,a\sigma_0)$
parameter space is consistent with both KKLT stability and the swampland criteria.

\item \textbf{Cosmological implications.}
The metastable de~Sitter vacua of Section~\ref{sec:moduli} are natural starting points for string
inflationary models.  Inflaton candidates include the volume modulus $\sigma$ (Kähler
moduli inflation), the axion $\theta$ (natural inflation), and brane positions (KKLMMT
inflation).  In each case the inflationary potential is a deformation of the scalar
potential studied in Section~6, and the slow-roll parameters $\epsilon$ and $\eta$
can be expressed in terms of the same quantities $\hat\xi$, $a$, $W_0$, and $A$ that
characterise the three regimes of Table~\ref{tab:regimes}.  Connecting the phase structure of
Section~6.3 with the inflationary observables $(n_s, r)$ would therefore place
direct observational constraints on the correction parameter $\hat\xi$ and on the
viability of KKLT as an inflationary framework.

\end{itemize}

%%%%%%%%%%%%%%%%%%%%%%%%

\end{document}